\documentclass[11pt,headings=big,numbers=noenddot,DIV=14,a4paper]{article}%

\pdfoutput=1

\usepackage[margin=1 in]{geometry}
\usepackage{comment}
\usepackage{graphicx}  
\usepackage{dcolumn}   
\usepackage{bm,relsize}        
\usepackage{amsfonts,amsmath,amssymb,amsthm,mathtools}
\usepackage{slashed} 
\usepackage[normalem]{ulem}
\usepackage{placeins}
\usepackage{lscape}
\usepackage{multirow}
\usepackage{float}
\usepackage{lipsum, color}
\usepackage[usenames,dvipsnames,svgnames,table]{xcolor}
\usepackage[sort&compress,numbers,merge]{natbib}
\usepackage{breakcites}
\renewcommand{\figurename}{Fig.}

\usepackage{lipsum, color}
\usepackage[usenames,dvipsnames,svgnames,table]{xcolor}
\definecolor{peera_col}{RGB}{240, 94, 28}
\definecolor{blue_col}{RGB}{0,92,175}
\definecolor{red_col}{RGB}{203,64,66}

\usepackage[linktoc=page,bookmarks=false,colorlinks=true,linkbordercolor=PineGreen,citebordercolor=PineGreen,urlbordercolor=PineGreen]{hyperref}
\hypersetup{
    colorlinks=true,
    linkcolor=red_col,
    filecolor=Mahogany,      
    urlcolor=blue_col,
    citecolor=blue_col,
}

\usepackage{chngcntr}
\counterwithin*{equation}{section}

\usepackage{fancyhdr}
\fancypagestyle{specialfooter}{%
  \fancyhf{}
  
  \fancyfoot[L]{{\noindent\rule{4cm}{0.5pt}\\
  \footnotesize E-mails: \textcolor{blue_col}{\texttt{anish.ghoshal@fuw.edu.pl}, \texttt{yann.gouttenoire@gmail.com}, \texttt{lucien.heurtier@durham.ac.uk}, \texttt{peera.simakachorn@ific.uv.es}}}}
}

\begin{document}

\color{Black}

\thispagestyle{specialfooter}

\begin{flushright}  IPPP/23/13
\end{flushright}

\bigskip
\begin{center}

{\LARGE \bf 
Primordial Black Hole Archaeology \vspace{9pt}\\ with Gravitational Waves from Cosmic Strings
}

\end{center}

\vspace{0.25cm}

\bigskip

\centerline{
{\large Anish Ghoshal}$^{a}$,
{\large Yann Gouttenoire}$^{b}$,
{\large Lucien Heurtier}$^{c}$,
{\large and Peera Simakachorn}$^{d}$
}

\bigskip
\begin{center}
{\it \small $^a$ Institute of Theoretical Physics, Faculty of Physics, University of Warsaw, \\ ul. Pasteura 5, 02-093 Warsaw, Poland}\\
{\it \small $^b$ School of Physics and Astronomy, Tel-Aviv University, Tel-Aviv 69978, Israel}\\
{\it \small $^c$ Institute for Particle Physics Phenomenology, Durham University, South Road, Durham, U.K.}\\
{\it \small $^d$ 
Instituto de F\'isica Corpuscular (IFIC), Universitat de Val\`{e}ncia-CSIC,\\C/ Catedrático José Beltrán 2, E-46980, Paterna, Spain}\\[0.75em]
\end{center}

\bigskip
\bigskip
\centerline{\bf \large Abstract}

\bigskip

Light primordial black holes (PBHs) with masses smaller than $10^9$ g ($10^{-24} M_\odot$) evaporate before the onset of Big-Bang nucleosynthesis, rendering their detection rather challenging. If efficiently produced, they may have dominated the universe energy density. 
We study how such an early matter-dominated era can be probed successfully using gravitational waves (GW) emitted by local and global cosmic strings. While previous studies showed that a matter era generates a single-step suppression of the GW spectrum, we instead find a \emph{double-step} suppression for local-string GW whose spectral shape provides information on the duration of the matter era. The presence of the two steps in the GW spectrum originates from GW being produced through two events separated in time: loop formation and loop decay, taking place either before or after the matter era. The second step -- called the \emph{knee} -- is a novel feature which is universal to any early matter-dominated era and is not only specific to PBHs. 
Detecting GWs from cosmic strings with LISA, ET, or BBO would set constraints on PBHs with masses between $10^6$ and $10^9$ g for local strings with tension $G\mu = 10^{-11}$, and PBHs masses between $10^4$ and $10^9$ g for global strings with symmetry-breaking scale $\eta = 10^{15}~\mathrm{GeV}$. Effects from the spin of PBHs are discussed. 

\clearpage

\tableofcontents
 \noindent\hrulefill

\section{Introduction}
\label{sec:intro}

\medskip

The precise measurement of the Cosmic Microwave Background (CMB) \cite{Planck:2018vyg} and the successful prediction of Big-Bang Nucleosynthesis (BBN) \cite{Kawasaki:1999na, Kawasaki:2000en, Hannestad:2004px} have determined the amount of radiation, baryons, dark matter and dark energy in our universe with extraordinary accuracy. Temperature anisotropies measured in the CMB superimposed on the perfectly smooth background suggest that our universe has started with an accelerated expansion called cosmic inflation~\cite{Guth:1980zm, Linde:1981mu, Albrecht:1982wi, Akrami:2018odb}. 
However, the cosmic history after inflation and before BBN -- above the plasma temperature of $T\gtrsim 1~\rm MeV$ -- is difficult to probe and is currently unconstrained by data. The standard assumption that the pre-BBN universe is radiation-dominated might be challenged by open problems in the Standard Model (SM) of particle physics, e.g., the origin of the matter-antimatter asymmetry, the origin of dark matter, the flavor puzzle, or the ultraviolet dynamics in the Higgs sector, see e.g. Ref.~\cite{Gouttenoire:2022gwi} for a review. New physics addressing those issues requires the introduction of new energy scales and new degrees of freedom, which sometimes generate deviations from the standard radiation domination era before the onset of BBN. Well-known examples are a long-lived heavy scalar field generating an early matter era~\cite{McDonald:1989jd,Moroi:1999zb, Visinelli:2009kt,Erickcek:2015jza,  Nelson:2018via,Cirelli:2018iax,Gouttenoire:2019rtn,Allahverdi:2020bys}, a fast-moving scalar field generating a kination era \cite{Spokoiny:1993kt,Joyce:1996cp,Peebles:1998qn,Poulin:2018dzj,Gouttenoire:2021jhk,Gouttenoire:2021wzu,Co:2021lkc,Ghoshal:2022ruy,Heurtier:2022rhf}, or a supercooled phase transition \cite{Guth:1980zk,Witten:1980ez,Creminelli:2001th,Randall:2006py,Konstandin:2011dr,vonHarling:2017yew,Baratella:2018pxi,Ghoshal:2020vud,Baldes:2020kam,Baldes:2021aph,Dasgupta:2022isg,Ferrer:2023uwz, Wong:2023qon}. In the present work, we are interested in the possibility that the early universe underwent a matter-dominated phase induced by a population of Primordial Black Holes (PBHs) evaporating before BBN.\footnote{Early eras of cosmology deviating from radiation domination may also arise when an extended particle physics sector or an extended distribution of PBHs decays/evaporate in the early universe, as studied in Refs.\cite{Barrow:1991dn,Dienes:2022zgd, Dienes:2021woi}.}

  Unlike astrophysical black holes, PBHs can have masses that range from $M_{\rm PBH} \lesssim 0.1~\mathrm{g}$ to $10^{23}~\mathrm{g}$ \cite{Carr:2020gox, Green:2020jor}. They originate from the collapse of primordial curvature over-densities \cite{Carr:1975qj}. Such primordial inhomogeneities can be sourced by cosmic inflation~\cite{Carr:1993aq,Ivanov:1994pa}, transitions to a metastable vacuum during inflation~\cite{Garriga:2015fdk,Deng:2016vzb,Deng:2017uwc,Kusenko:2020pcg}, supercooled phase transitions~\cite{Sato:1981bf,Maeda:1981gw,Sato:1981gv,Kodama:1981gu,Kodama:1982sf,Hsu:1990fg,Liu:2021svg,Kawana:2022olo,Gouttenoire:2023naa}, bubble collisions \cite{Hawking:1982ga,Moss:1994iq,Moss:1994pi, Khlopov:1998nm}, matter squeezing by bubble walls \cite{Crawford:1982yz,Gross:2021qgx,Baker:2021sno,Kawana:2021tde}, collapse of scalar condensate \cite{Dolgov:1992pu,Dolgov:2008wu,Kitajima:2020kig,Kasai:2022vhq,Martin:2019nuw,Martin:2020fgl}, of domain walls~\cite{Rubin:2000dq, Vachaspati:2017hjw,Ferrer:2018uiu,Gelmini:2022nim,Gelmini:2023ngs}, or of cosmic-string loops \cite{Hawking:1987bn,Polnarev:1988dh,Fort:1993zb,
Garriga:1993gj,Caldwell:1995fu,MacGibbon:1997pu,Jenkins:2020ctp,Blanco-Pillado:2021klh}. PBHs lighter than $10^{15}~\mathrm{g}$ would have already evaporated by now~\cite{Hawking:1974rv, Hawking:1974sw}. The ones in the range $10^9~\mathrm{g}\lesssim M_{\rm PBH}\lesssim 10^{15}~\mathrm{g}$ would have evaporated after the onset of BBN and are strongly constrained by cosmological observations \cite{Keith:2020jww,Poulin:2016anj}. However, PBHs lighter than $10^9~\mathrm{g}$ evaporate in the pre-BBN universe, and their observational signatures are considerably limited.

The detection of gravitational waves (GWs) from astrophysical sources by the LIGO-Virgo collaboration in 2015~\cite{Abbott:2017xzu} opened the door to GW astronomy. Upcoming upgrades of LIGO-Virgo \cite{Aasi:2014mqd} and proposed future detectors such as LISA~\cite{Audley:2017drz}, BBO-DECIGO~\cite{Yagi:2011wg}, the 
Einstein Telescope~(ET)~\cite{Punturo:2010zz,Hild:2010id},
and Cosmic Explorer~(CE)~\cite{Evans:2016mbw} will open up a new observation window of the early universe. Unlike photons, primordial GWs that were emitted in the early Universe can propagate freely throughout cosmic history and therefore would constitute ideal messengers of our Universe history \cite{Allen:1996vm, Caprini:2018mtu,Simakachorn:2022yjy}.
One of the pinnacles for determining the pre-BBN cosmic history of the universe would be the detection of GW sourced by a network of cosmic strings. Cosmic strings are one-dimensional objects produced by the spontaneous breaking of a $U(1)$ symmetry in the early universe~\cite{Nielsen:1973cs, Kibble:1976sj}, the pure Yang-Mills theory \cite{Yamada:2022imq, Yamada:2022aax}, or fundamental objects in superstring theory~\cite{Copeland:2003bj, Dvali:2003zj, Polchinski:2004ia, Jackson:2004zg, Tye:2005fn}. The crucial peculiarity of cosmic strings is that they are long-standing GW sources \cite{Vilenkin:1981bx,Vachaspati:1984gt,  Hindmarsh:1994re, Vilenkin:2000jqa}. Once the network of cosmic strings is produced, it occupies a constant fraction of the total energy density of the universe, a property known as the scaling regime \cite{Albrecht:1984xv, Bennett:1987vf, Allen:1990tv,Martins:2000cs, Figueroa:2012kw, Martins:2016ois}. An important consequence is that GW emissions occur during most of the universe history.  This generates a GW spectrum spanning many orders of magnitude in frequencies. A measurement of the GW spectrum from high to lower frequencies would allow to determine the universe expansion rate from early to later times \cite{Cui:2017ufi, Cui:2018rwi,  Ramberg:2019dgi, Gouttenoire:2019kij,Gouttenoire:2019rtn,Blasi:2020mfx, Datta:2020bht, Samanta:2021mdm, Borah:2022iym}. Instead, the use of short-lived GW sources like the ones from first-order phase transitions can only probe the cosmic history within a small window of time around when the source is active \cite{Barenboim:2016mjm,Hook:2020phx,Ellis:2020nnr,Gouttenoire:2021jhk,Domenech:2020kqm}.
Another long-standing GW source that could bring information about the equation of state of the pre-BBN universe is tensor modes sourced during primordial inflation~\cite{Giovannini:1998bp, Riazuelo:2000fc, Sahni:2001qp, Tashiro:2003qp,Boyle:2007zx}. The BICEP-Keck bound on the tensor-to-scalar ratio $r\lesssim 0.036$ \cite{BICEP:2021xfz} constrains the inflation scale to $H_{\rm inf}\lesssim 3\times 10^{13}~\rm GeV$ which prevents the detectability of tensor modes by GW detectors in a foreseeable future~\cite{ Smith:2005mm, Ananda:2006af,Lasky:2015lej,Guzzetti:2016mkm}. However, intermediate cosmological eras can imprint signatures on such GW spectrum \cite{DEramo:2019tit, Bernal:2020ywq,Gouttenoire:2021jhk,Gouttenoire:2021wzu,Dunsky:2021tih,Berbig:2023yyy}.

\begin{figure}[!ht]
\centering
\includegraphics[width=0.485\textwidth, scale=1]{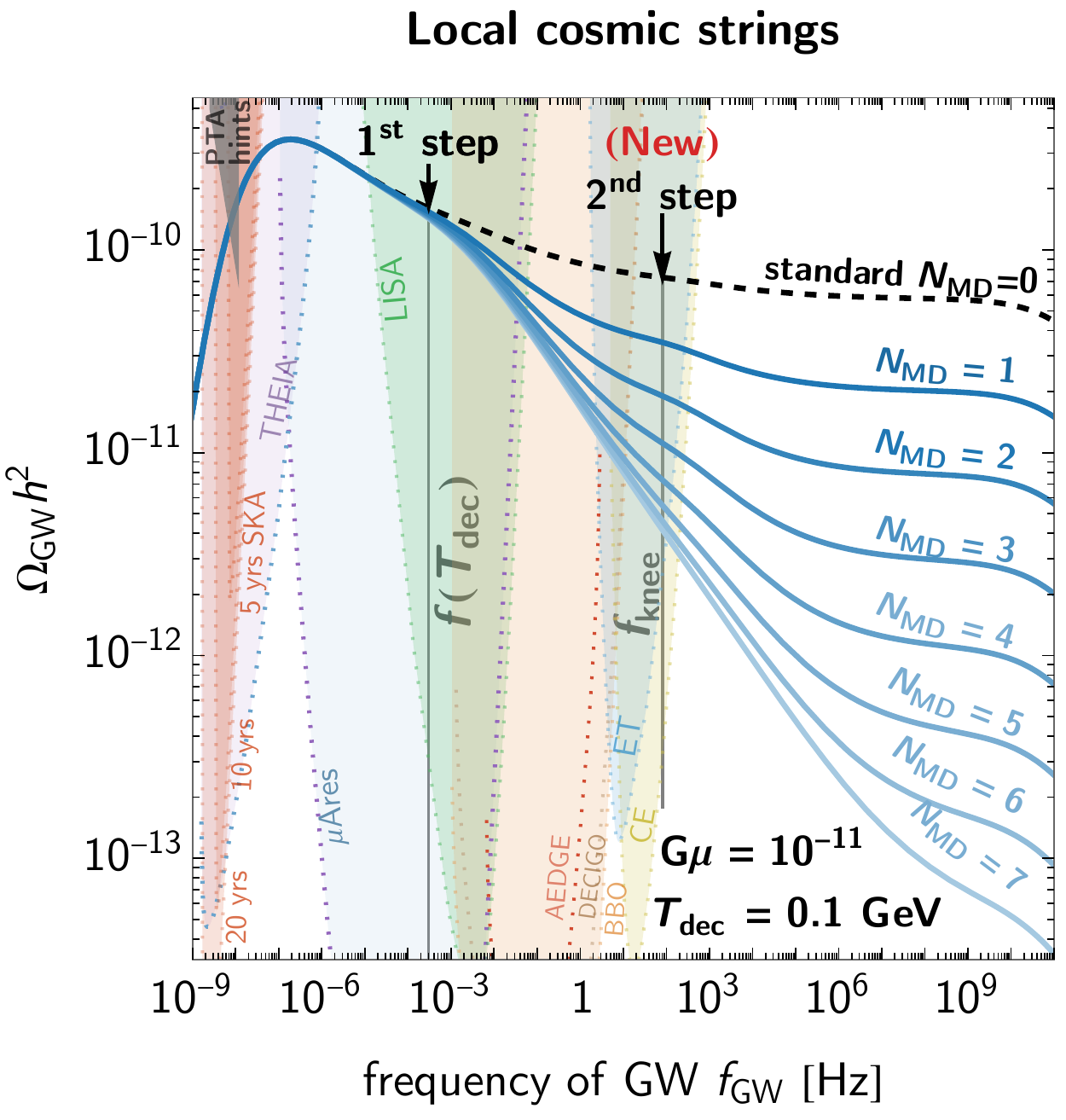}
\hfill
\includegraphics[width=0.4965\textwidth, scale=1]{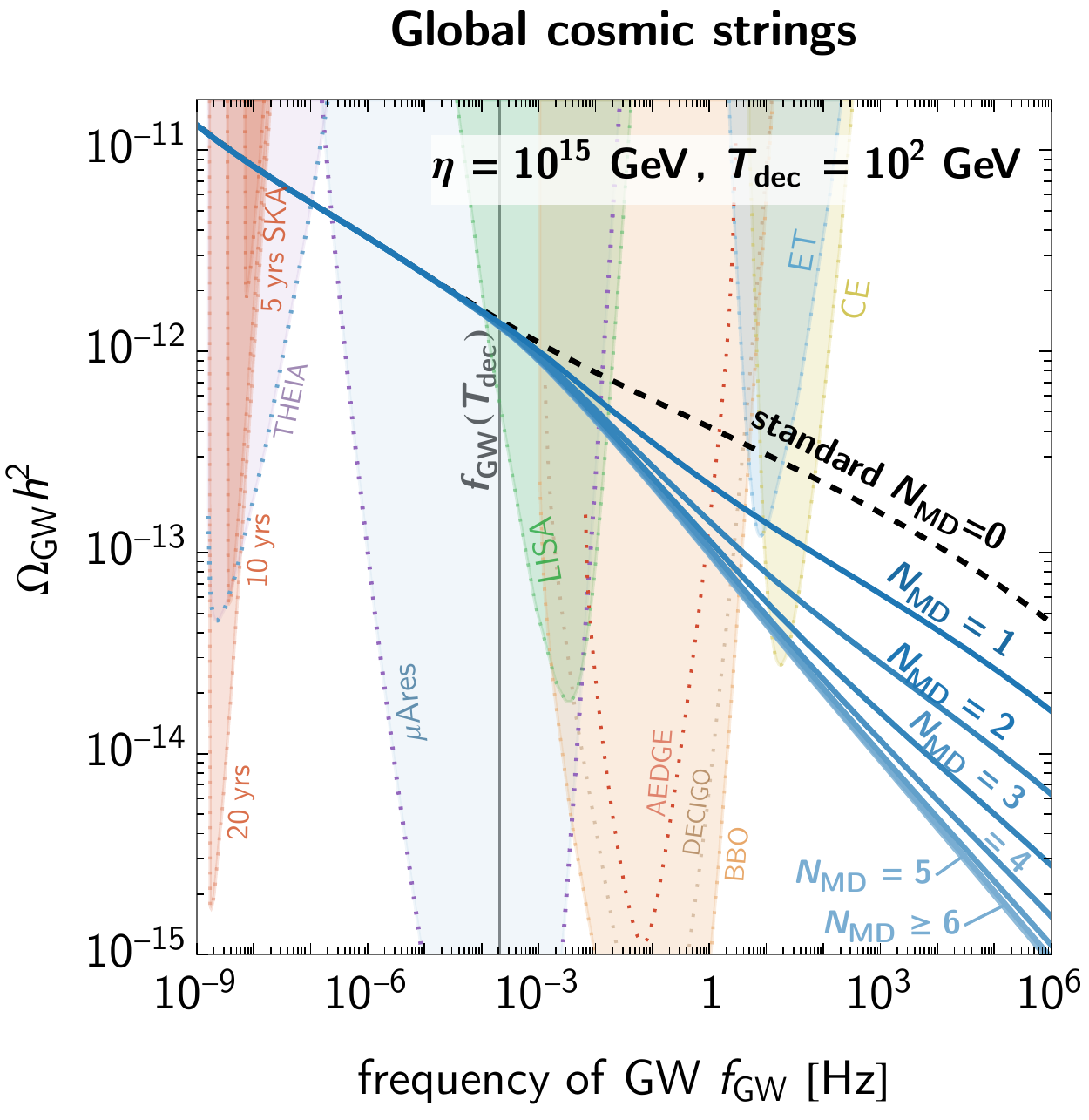}\\[-1em]
\caption{\textit{ \small We assume an early matter domination (e.g., PBH domination era) lasting for $N_{\rm MD}$ e-folds of Hubble expansion and ending at temperature $T_{\rm dec}$. It impacts the GW spectra of local and global cosmic strings through a double-step (\textbf{left}) and a single-step (\textbf{right}) suppression, respectively. Those features can be characterized by three frequencies: the low-frequency turning point, the high-frequency plateau, and the newly-found knee (the second step in local strings GW spectrum)}, see Sec.~\ref{sec:matter_era}.}
\label{fig:spectrum_strings}
\end{figure}

In this work, we explore how a period of early matter domination that originates from a large production of PBHs in the early Universe could leave observable imprints in the spectrum of GWs generated by a network of cosmic strings. We do not assume any connection between the sector producing PBHs and the one producing cosmic strings. \textit{The paper is organized as follows:} in Sec.~\ref{sec:GW_CS}, we review the calculation of the GW spectrum from cosmic strings, both in the local and global cases,  as illustrated in Fig.~\ref{fig:spectrum_strings}. Readers familiar with cosmic strings can jump directly to Sec.~\ref{sec:matter_era} where we study the impact of a matter-dominated era. The main smoking-gun GW signatures are summarized in Fig.~\ref{fig:spectral_feature}, including the \emph{double-step} feature and its associated \emph{knee}, which are studied for the first time in this work. Finally, in Sec.~\ref{sec:PBH_era}, we use those findings to constrain the abundance and mass of primordial black holes formed with a monochromatic distribution, see Fig.~\ref{fig:local_pbh_plane} and  \ref{fig:global_pbh_plane} for local and global strings, respectively. We would like to emphasize that the PBHs considered in this work evaporated well before BBN. Hence the constraints are shown as the fraction of energy density $\beta$ in PBH at formation, instead of today PBH fraction $\Omega_{\rm PBH}$. We discuss these results and conclude in Sec.~\ref{sec:conclusion}.

\section{GW from Cosmic Strings}
\label{sec:GW_CS}

A cosmological phase transition associated with the spontaneous breaking of $U(1)$ symmetry leads to the formation of a network of topological defects called \emph{cosmic strings} (CS). The $U(1)$ symmetry can be either \emph{local} or \emph{global}. We refer to Ref.~\cite{Kibble:1976sj} for the original article, Ref.~\cite{Vilenkin:2000jqa} for a textbook, and Refs.~\cite{Sousa:2014gka, Gouttenoire:2019kij, Auclair:2019wcv, Gouttenoire:2022gwi, Simakachorn:2022yjy} for reviews of their GW emission.

Cosmic strings are field configurations at the top of the $U(1)$-breaking Mexican hat potential. The field energy is localized within a core of inverse symmetry-breaking scale size, much smaller than the cosmological horizon. This motivates the \emph{Nambu-Goto} approximation which describes CS as infinitely thin classical objects with energy per unit length $\mu$,
\begin{equation}
\mu = 2\pi n \, \eta^2 \times \begin{cases}
1 ~ ~ ~ & {\rm for ~ local ~ strings},\\
\log(\eta t)  ~ ~ ~ & {\rm for ~ global ~ strings},
\end{cases}
\label{string_tension}
\end{equation} 
where $\eta$ is the vacuum expectation value of the scalar field constituting CS, and $n$ is the winding number taken to be $n=1$ since it is the only stable configuration \cite{Laguna:1989hn}.  For global strings, the logarithmic divergence arises due to the massless Goldstone mode leading to the presence of long-range gradient energy \cite{Vilenkin:2000jqa}. The CS network forms at the temperature of the $U(1)$-breaking phase transition,
\begin{align}
T_\textrm{form} &\simeq \eta \simeq 10^{11} \textrm{ GeV}\left(\frac{G\mu}{10^{-15}}\right)^{1/2},
\label{string_formation_cutoff}
\end{align}
where the second equality refers to local strings.

\subsection{Evolution of the network}

In general, one should expect the motion of cosmic strings to be initially frozen due to thermal friction, characterized by the frictional damping length $l_f = \mu/(\beta T^3)$ where $\beta =\mathcal{O}(1)$ represents the strength of particle-string interactions \cite{Vilenkin:1991zk,Martins:1995tg,Martins:1996jp}. At that time the energy density of frozen CS evolves under Hubble expansion as $\rho_{\rm CS} \propto a^{-2}$ where $a$ is the scale factor \cite{Vilenkin:2000jqa}, and would quickly dominate the energy density of the universe.
Effects from friction become negligible when the damping length $l_f$ becomes larger than the cosmic horizon $t$, i.e., when
\begin{align}
\label{eq:Tfric_def}
T ~\lesssim ~T_{\rm fric} = \frac{4 \times 10^9{~\rm GeV~}}{\beta}\left( \frac{g_*}{100} \right)^{1/2} \left(\frac{G\mu}{10^{-11}}\right).
\end{align}
Below that temperature, CS start to move freely under the work of their own tension, they reach velocities of order $\mathcal{O}(0.1)$ and start to interact with each other \cite{Vilenkin:2000jqa}. The intersection of straight strings forms loops that decay by radiating particles and GW. Loop formation acts as a loss mechanism for the infinitely long-string network. Out of those two antagonist effects, Hubble expansion and loop formation, which respectively increase and decrease the fractional CS abundance $\Omega_{\rm CS} \equiv \rho_{\rm CS} / \rho_{\rm tot}$, the CS network reaches a stable state called the \emph{scaling regime} where $\Omega_{\rm CS}$ remains constant over time \cite{Albrecht:1984xv, Bennett:1987vf, Allen:1990tv,Martins:2000cs, Figueroa:2012kw, Martins:2016ois}. In the scaling regime, CS redshifts the same way as the background, e.g., $\rho_{\rm CS} \propto a^{-4}$ during radiation-domination (RD) and  $\rho_{\rm CS} \propto a^{-3}$ during matter-domination (MD). One can say that the equation of state of the CS network tracks the equation of state of the expanding background.

The GW signal from CS is dominated by emission from loops \cite{Vachaspati:1984gt, Allen:1991bk,Vilenkin:2000jqa}.
Loops of strings are constantly being produced as long strings self-intersect. 
Local string loops decay into GW after $0.001/G\mu \gg 1$ Hubble time, while global string loops decay into Goldstone modes in less than one Hubble time, cf. Eq.~\eqref{eq:tilde_t_M}.

\begin{itemize}
\item \textbf{Step 1: Loop Formation.}
We denote $\alpha$ the size of loops in units of the cosmic horizon $t_i$ when they form.
The rate of loop formation rate can be written \cite{Vilenkin:2000jqa}
\begin{align}
\frac{dn_{\rm loop}}{dt_i} = 0.1 \frac{C_{\rm eff}(t_i)}{\alpha t_i^4},
\end{align}
where factor 0.1 indicates that 90\% of the loop population is small and highly-boosted loops that are red-shifted away and do not contribute substantially to the GW signal \cite{Blanco-Pillado:2013qja}. We have introduced the loop formation efficiency factor $C_{\rm eff}$, which for local strings reaches the asymptotic value $C_{\rm eff} \simeq 0.39$, $5.4$, $29.6$ during matter-domination (MD), radiation-domination (RD), and kination eras, respectively \cite{Gouttenoire:2019kij}. 
For global strings, the long strings lose energy via Goldstone emission on top of loop formation, which logarithmically suppresses the loop production efficiency. At the analytical level, we take $C_{\rm eff} \sim \mathcal{O}(1)$ for all cosmological eras \cite{Chang:2019mza, Gouttenoire:2019kij, Chang:2021afa}. 
However, for the plots and analysis of this paper, we solve $C_{\rm eff}(t)$ as a solution of the velocity-dependent one-scale (VOS) equations governing the string network evolution. As studied thoroughly in Ref.~\cite{Sousa:2013aaa,Gouttenoire:2019kij}, the VOS evolution captures the inertia of the network during a change in the equation of state of the universe. Instead, approximating $C_{\rm eff}$ using a piecewise constant function across different eras overestimates the value of the turning-point frequency in Eq.~\eqref{turning_point_CS} by more than one order of magnitude \cite{Gouttenoire:2019kij}.

\item \textbf{Step 2: GW Emission from Loops.}
Numerical simulations \cite{Blanco-Pillado:2013qja} have shown that the GW spectrum is dominantly produced by loops with the largest size, corresponding to $10\%$ of the horizon. We account for this result by choosing a monochromatic loop size probability distribution 
\begin{align}
\mathcal{P}_{\rm loop}(\alpha) = \delta(\alpha - 0.1).
\end{align}
After their formation, $\tilde t > t_i$, loops of length $l(\tilde{t})$ oscillate and radiate a discrete spectrum of GWs with frequencies given by  
\begin{equation}
\label{eq:emitted_frequency}
\tilde{f} = 2k/l(\tilde{t}), \qquad  k \in \mathbb{Z}^+.
\end{equation}
The frequency today is given by $f=\tilde{f} a(\tilde{t})/a_0$.
The GW emission power by a loop is independent of its size, which is a remarkable result of the quadrupole formula \cite{Gouttenoire:2019kij}. For each Fourier mode $k$, it is given by
\begin{align}
P_{\rm GW}^{(k)} = \Gamma^{(k)} G\mu^2, \qquad \text{with} \quad \Gamma^{(k)} = \frac{\Gamma k^{-{\delta}}}{\sum_{p=1}^\infty p^{-\delta}},
\label{eq:power_emisssion_GW_strings}
\end{align}
and $\Gamma = 50$ for local \cite{Blanco-Pillado:2017oxo} and global strings \cite{Gorghetto:2021fsn}. 
The index $\delta$ depends on whether high Fourier modes are dominated by cusps ($\delta = 4/3$), kinks ($\delta =5/3$), and kink-kink collisions  $(\delta=2)$ \cite{Olmez:2010bi}. This choice does not impact much the amplitude of the GW spectrum in the deep  RD and MD era, but it substantially impacts the slope around the transition between early MD and RD, see Sec.~\ref{sec:spectral_index}. In this work, we assume the small-scale structure to be dominated by cusps $\delta = 4/3$.
While loops continuously lose energy as they emit either GWs or Goldstone modes, their length $l$ shrinks as 
\begin{align}
l(\tilde{t}) =  \alpha t_i - (\Gamma G \mu + \kappa) (\tilde{t} - t_i). \label{eq:length_shrink}
\end{align}
where $\Gamma G \mu$ and $\kappa$ are the shrinking rates due to GW and particle emissions, respectively.
Local-string loops dominantly decay via GW emission ($\kappa = 0$), while global-string loops dominantly decay into Goldstone modes with $\kappa = \Gamma_{\rm Gold}/2 \pi \log(\eta t) \gg \Gamma G\mu$ where $\Gamma_{\rm Gold} \simeq 65 $ \cite{Vilenkin:1986ku}. 
\end{itemize}

\subsection{GW spectrum}
From right to left, we write in chronological order the different involved processes leading to the final expression for the spectral energy density of GWs from CS, defined as $\Omega_{\rm GW}\equiv \frac{1}{\rho_c}\frac{d\rho_{\rm GW}}{d\ln{f}}$,
\begin{align}
\Omega_{\rm GW}(f) &= \sum_k \frac{1}{\rho_c} \int_{t_{\rm osc}}^{t_0} d\tilde{t} \int_0^1 d\alpha \,\Theta\left[t_i - \frac{l_*}{\alpha}\right]\cdot \Theta[t_i - t_{\rm osc}]\cdot\left[\frac{a(\tilde{t})}{a(t_0)} \right]^4 \cdot P_{\rm GW}^{(k)} \times \nonumber\\
&\hspace{15em} \times\left[\frac{a(t_i)}{a(\tilde{t})} \right]^3 \cdot \mathcal{P}_{\rm loop}(\alpha)\cdot \frac{dt_i}{df}\cdot \frac{d n_{\rm loop}}{d t_i}. \label{eq:master_eq_pedagogical}
\end{align}
First, loops are formed at a rate $dn_{\rm loop}/dt_i$ (note the chain rule $dt_i/df$) with a distribution of size $\mathcal{P}_{\rm loop}(\alpha)$. They redshift like $a^{-3}$ before radiating GW with power $P_{\rm GW}^{(k)}$ which subsequently dilute as $a^{-4}$. We add two Heaviside functions to cut off the GW spectrum above some characteristic frequencies discussed below. Finally, we integrate over all loop sizes $\alpha$ and GW emission times $\tilde{t}$ and sum over all Fourier modes $k$.  

The first Heaviside function $\Theta(t_i- l_*/\alpha)$ eliminates loops smaller than a critical length $l_*$ below which massive particle production is the main decay channel \cite{Auclair:2019jip,Gouttenoire:2019kij}.
The second Heaviside function $\Theta(t_i-t_{\rm osc})$ with $t_{\rm osc}=\text{Max}\left[t_{\rm form},\,t_{\rm fric} \right]$ gets rid of loops which would have formed before the formation of the network, cf. Eq.~\eqref{string_formation_cutoff}, or which form in the friction-dominated epoch, cf. Eq.~\eqref{eq:Tfric_def}.
As shown in \cite{Gouttenoire:2019kij}, these high-frequency cut-offs lie at frequencies higher than the windows of current and future GW interferometers. 
After this short introduction, we simplifies Eq.~\eqref{eq:master_eq_pedagogical} to the handy form,
	\begin{align}
\Omega_{\rm GW}(f) &=\sum_k\frac{1}{\rho_c}\cdot\frac{2k}{f}\cdot\frac{(0.1) \,\Gamma^{(k)}G\mu^2}{\alpha(\alpha+\Gamma G \mu + \kappa)} \times \nonumber\\
&\hspace{3em} \times \int^{t_0}_{t_{\rm osc}}d\tilde{t} \, \frac{C_{\rm{eff}}(t_i)}{t_i^4}\left[\frac{a(\tilde{t})}{a(t_0)}\right]^5\left[\frac{a(t_i)}{a(\tilde{t})}\right]^3\Theta\left(t_i-\frac{l_*}{\alpha}\right)\Theta(t_i-t_{\rm osc}).
	\label{eq:master_eq_ready_to_use}
	\end{align}
This formula can be used for both local and global strings after applying Eqs.~\eqref{eq:emitted_frequency} and \eqref{eq:length_shrink} and choosing an appropriated value of $\kappa$.	
The resulting GW spectra assuming the standard $\Lambda$CDM Universe are illustrated by the black dashed lines in Fig.~\ref{fig:spectrum_strings}.
For loops formed and emitted during radiation-domination epoch, the GW spectrum emitted by local strings is nearly flat with an amplitude of order\footnote{Keeping all the parameters, we get  \cite{Gouttenoire:2019kij} 
\begin{equation}
    \Omega_{\rm std}^{\rm CS} h^2 \simeq 1.5 \pi\Omega_r h^2 \, \mathcal{G}\,C_{\rm eff}^{\rm rad} \left(\frac{\alpha G \mu}{\Gamma}\right)^{1/2}, \qquad \text{(local)}.
\end{equation}} 
\begin{equation}
    \Omega_{\rm std}^{\rm CS} h^2 \simeq \Omega_r h^2 \mathcal{G}(\tilde{T}_{\rm M})\left( \frac{\eta}{M_{\rm pl}} \right),
\end{equation}
where $\Omega_{r}h^2 \simeq 4.2 \times 10^{-5}$ \cite{ParticleDataGroup:2020ssz} is the radiation density today and $\tilde{T}_{\rm M}$ is the temperature at time $\tilde{t}_{\rm M}$ of maximum emission defined in Eq.~\eqref{eq:tilde_t_M}.
Deviations from flatness arise due to a change in the number of relativistic degrees of freedom captured in \cite{Gouttenoire:2019kij} 
\begin{align}
\label{eq:Delta_R}
\mathcal{G}(T) \equiv \left( \frac{g_*(T)}{g_*(T_0)}\right)\left(\frac{g_*s(T_0)}{g_{*s}(T)} \right)^{4/3} = ~0.39 \left(\frac{106.75}{g_*(T)} \right)^{1/3}.
\end{align}
In comparison, the GW spectrum from global strings is suppressed by $(\eta/M_{\rm pl})^3$ due to the shorter loop lifetime and enhanced by a $\log^3$ factor due to the larger string tension\footnote{The GW spectrum of global strings in radiation-dominated universe reads \cite{Gouttenoire:2019kij}
\begin{equation}
\Omega_{\rm std}^{\rm CS} h^2 \simeq 9\Omega_r h^2 \, \mathcal{G}(\tilde{T}_{\rm M})\,C_{\rm eff}^{\rm rad} \left(\frac{\Gamma}{\Gamma_{\rm gold}}\right) \left(\frac{\eta}{M_{\rm pl}}\right)^{4} \log^3\left( \eta \tilde{t}_{\rm M}\right),\qquad \text{(global)}.
\end{equation}
 }
\begin{equation}
    \Omega_{\rm std}^{\rm CS} h^2 \sim \Omega_r h^2 \mathcal{G}(\tilde{T}_{\rm M})\left( \frac{\eta}{M_{\rm pl}} \right)^4 \log^3\left( \eta \tilde{t}_{\rm M}\right).
\end{equation}
We refer to Ref.~\cite{Gouttenoire:2019kij} for a detailed discussion about the differences between GW from local and global strings.
The impact of a non-standard cosmological evolution, such as a PBH-dominated universe, is studied in Sec.~\ref{sec:PBH_era}.

\subsection{Temperature-frequency relation}
\label{sec:freq_T_relation}
The dominant emission in GW arises at the end of the loops lifetime when the loop size in Eq.~\eqref{eq:length_shrink} is half its initial size $l(\tilde{t}_{\rm M})=\alpha t_i/2$ \cite{Cui:2018rwi, Gouttenoire:2019kij}
\begin{equation}
\label{eq:tilde_t_M}
    \tilde{t}_{\rm M} = \frac{\alpha/2 + \Gamma G\mu + \kappa}{\Gamma G\mu + \kappa}t_i~\simeq~  \begin{cases}
\frac{\alpha}{2 \Gamma G \mu} t_i ~ ~ &\mathrm{for} ~ \kappa = 0, ~ \alpha \gg \Gamma G \mu ~ \qquad \mathrm{(local)},\\
t_i ~ ~ &\mathrm{for} ~ \kappa \gg \alpha \gg \Gamma G \mu ~ \qquad~~~ \mathrm{(global)}.\\
\end{cases}
\end{equation}
The global-string loops decay fast after their production, while the local-string loops live much longer. The frequency today $f$ is related to the emitted frequency $\tilde{f}$ by $f=\tilde{f}\times a(\tilde{t})/a_0$. The emitted frequency $\tilde{f}$ is related to the loop length $l(\tilde{t}_{\rm M})=\alpha t_i/2$ by Eq.~\eqref{eq:emitted_frequency}.
Assuming RD followed by the standard $\Lambda\rm CDM$ evolution after loop formation, we obtain the relation between the GW frequency today and the temperature when the loop dominantly sourcing this frequency mode is produced 
\begin{align}
f_\Delta ~ \simeq ~ \begin{cases}
(2 \times 10^{-3} ~ \mathrm{Hz}) \left(\frac{0.1 \times 50 \times 10^{-11} }{\alpha \times \Gamma G\mu}\right)^{1/2}\left(\frac{T_\Delta}{\rm GeV}\right) \left[\frac{g_*(T_\Delta)}{g_*(T_0)}\right]^{\frac{1}{4}} ~ ~ ~ &\mathrm{(local ~ strings)},\\[1em]
(4.7 \times 10^{-6} ~ \mathrm{Hz}) \left(\frac{0.1}{\alpha}\right) \left(\frac{T_\Delta}{\rm GeV}\right) \left[\frac{g_*(T_\Delta)}{g_*(T_0)}\right]^{\frac{1}{4}} ~ ~ ~ &\mathrm{(global ~ strings)},
\end{cases}
\label{turning_point_CS}
\end{align}
We have multiplied the numerical-fitted factors\footnote{Since these factors depend on how fast the string network evolves, the local-string network reaches the scaling regime slower than the global string. The reason is that global strings lose energy more efficiently than local strings.} of $\approx 0.03$ and $\approx 0.2$ for local and global strings, respectively, to account for VOS evolution \cite{Gouttenoire:2019kij}.

We now use Eq.~\eqref{turning_point_CS} to compute the frequencies beyond which the GW spectrum is cut off due to network formation, friction, and particle production. Those cut-offs are away from the reach of current and planned GW interferometers. This has to be contrasted with the frequency of the turning point caused by the presence of an early matter-domination era -- and in particular, a PBH-dominated era -- which can lie within the detectable windows. 

\textbf{Formation cut-off.}--
No GW can be produced before the cosmic strings network is formed.
Plugging the temperature at which the $U(1)$ symmetry is spontaneously broken in Eq.~\eqref{string_formation_cutoff} into Eq.~\eqref{turning_point_CS} leads to the high-frequency cut-off 
\begin{align}
f_\textrm{form} ~ \simeq \begin{cases}
206 ~  {\rm GHz}  \left(\frac{0.1 \times 50}{\alpha \Gamma}\right)^{\frac{1}{2}} \left[\frac{g_*(T_{\rm form})}{g_*(T_0)}\right]^{\frac{1}{4}}~  ~ &{\rm for ~ local},\\[0.5em]
0.47 ~  {\rm GHz} \left(\frac{T_{\rm form}}{10^{14} ~ {\rm GeV}}\right) \left(\frac{0.1}{\alpha}\right) \left[\frac{g_*(T_{\rm form})}{g_*(T_0)}\right]^{\frac{1}{4}}~  ~ &{\rm for ~ global},
\end{cases} 
\label{eq:freq_string_formation_cutoff}
\end{align}
where $T_0$ denotes the temperature of the Universe today. We recall that $\alpha$ is the string size at formation in a unit of the cosmic horizon and $\Gamma$ is the GW emission efficiency of a string loop.

\textbf{Friction cut-off.}--
GW emission starts when effects from friction become negligible and cosmic strings start oscillating at relativistic speed.
Plugging Eq.~\eqref{eq:Tfric_def} into Eq.~\eqref{eq:freq_string_formation_cutoff}, we obtain the high-frequency cut-off due to loop  motion being frozen at earlier times
\begin{align}
f_\textrm{fric} ~ \simeq  \begin{cases}
\frac{5 \times 10^7{~\rm Hz} }{\beta} \left(\frac{0.1 \times 50\times G\mu}{\alpha \Gamma\times 10^{-11}}\right)^{\frac{1}{2}} \left[\frac{g_*(T_{\rm fric})}{g_*(T_0)}\right]^{\frac{1}{4}} \left[\frac{g_*(T_{\rm fric})}{100}\right]^{\frac{1}{2}}~  ~ &{\rm for ~ local},\\[0.5em]
\frac{5.6 \times 10^3 ~ {\rm Hz}}{\beta} \log(\eta \tilde{t}_{\rm M}) \left(\frac{\eta}{10^{14}{~\rm GeV}}\right)^2 \left(\frac{0.1}{\alpha}\right) \left[\frac{g_*(T_{\rm fric})}{g_*(T_0)}\right]^{\frac{1}{4}}~  ~ &{\rm for ~ global}.
\end{cases} 
\label{eq:freq_string_friction_cutoff}
\end{align}

\textbf{Massive particle production cut-offs.}--
As previously discussed, global strings exhibit suppressed GW emission due to the efficient production of massless Goldstone modes. On the other hand, particles in the gauge sector of local strings are massive, resulting in significant suppression of their production rate \cite{Moore:1998gp, Olum:1999sg, Moore:2001px}. The emission of massive particles occurs when Fourier modes surpass the mass gap, which happens during cusps or kink-kink collisions. It has been found that the power emitted in massive particles only exceeds the power emitted in GWs for loops smaller than the critical length \cite{Auclair:2019jip,Gouttenoire:2019kij}\footnote{See \cite{Vincent:1997cx, Hindmarsh:2008dw, Hindmarsh:2017qff} for different findings.}
\begin{align}
\label{eq:critical_length_particle}
l_{\rm part}=\beta_m\frac{\mu^{-1/2}}{(\Gamma G\mu)^m},
\end{align}
where $m=1$ or $2$ for loops kink-dominated \cite{Matsunami:2019fss} or cusp-dominated \cite{Blanco-Pillado:1998tyu, Olum:1998ag,Blanco-Pillado:2015ana}, respectively, and $\beta_m\sim\mathcal{O}(1)$.
Loops with lengths smaller than $l_{\rm part}$ should be subtracted when computing the SGWB. Plugging Eq.~\eqref{eq:critical_length_particle} into Eqs.~\eqref{eq:emitted_frequency} and \eqref{turning_point_CS} gives the frequency cut-offs
\begin{equation}
f_{\rm part} \simeq
\begin{cases}
\text{(0.5 GHz)} ~  \sqrt{\dfrac{1}{\beta_c}} \,\left( \dfrac{G\mu}{10^{-15}} \right)^{1/4}&\textrm{\hspace{0.5em}for kinks},\\[1em]
\text{(100 GHz)} ~ \sqrt{\dfrac{1}{\beta_k}} \left( \dfrac{G\mu}{10^{-15}}  \right)^{3/4}&\textrm{\hspace{0.5em}for cusps}.\\
\end{cases}
\label{UVcutoff_f_app}
\end{equation}

\subsection{Current constraints on cosmic strings}

\textbf{Local strings.}--An SGWB might have been detected by pulsar timing arrays NANOGrav \cite{NANOGrav:2020bcs}, EPTA \cite{Chen:2021rqp}, PPTA \cite{Goncharov:2021oub}, and IPTA \cite{Antoniadis:2022pcn}. The signal from NANOGrav can be interpreted as an SGWB from cosmic strings with tension $G\mu \sim 8 \times 10^{-11}$ \cite{Ellis:2020ena,Blasi:2020mfx}. This interpretation is less favored by other PTAs which prefer a harder spectrum \cite{Bian:2022tju,Chen:2022azo}. Analysis of the PPTA data set provides the upper bound $G\mu \leq 5 \times 10^{-10}$ \cite{Chen:2022azo}. 

Other signatures from Nambu-Goto strings result from the static gravitational field around the string. This can induce gravitational lensing and temperature anisotropies in the CMB. The resulting constraint $G\mu \lesssim \, \text{few} \times 10^{-7}$, e.g., \cite{Christiansen:2010zi,Planck:2013mgr}, is however much looser than the one from GW production. However, a recent study has shown that the strong gravitational lensing of the fast radio bursts could probe down to $G\mu \sim 10^{-9}$ with future radio telescopes \cite{Xiao:2022hkl}.

\textbf{Global strings.}--
Global cosmic strings efficiently produce massless Goldstone particles that contribute to the number $N_{\rm eff}$ of effective relativistic degrees of freedom. The precise constraint relies on the abundance of Goldstone particles from strings which is still debatable.\footnote{Recent studies \cite{Hindmarsh:2019csc,Hindmarsh:2021vih,Buschmann:2019icd,Buschmann:2021sdq} propose that the Goldstone energy spectrum from strings is scale-invariant, while other studies \cite{Gorghetto:2018myk,Gorghetto:2020qws,Gorghetto:2021fsn} suggest a slightly infrared-dominated spectrum, which leads to the production of more Goldstone particles.} We quote the upper bound $\eta \lesssim 3.5 \cdot 10^{15} \, {\rm GeV}$ derived in Ref.~\cite{Chang:2021afa} and refer to Refs.~\cite{Gorghetto:2021fsn,Dror:2021nyr} for slightly tighter bounds.

The absence of B-mode polarization in the CMB provides another constraint on global strings. Assuming instantaneous reheating and only SM degrees of freedom, the upper limit on the inflationary Hubble parameter $H_{\rm inf}\lesssim 3\times 10^{13}~\rm GeV$ \cite{BICEP:2021xfz} translates to the maximum temperature of the universe $T_{\rm max} \lesssim 4 \times 10^{15}$~GeV. For the string network to form, the string scale $\eta$ must be smaller than the maximum temperature $\eta \lesssim 4 \times 10^{15}$~GeV, up to $\mathcal{O}(1)$ model-dependent parameters.

For $\eta \gtrsim 10^{15} \, {\rm GeV}$, GW from global strings extend to $f \lesssim 10^{-14} \, {\rm Hz}$ which could leave signature in CMB polarization experiments, e.g. Ref.~\cite{BICEP:2021xfz}.  Nonetheless, GW in this frequency range is produced after photon decoupling, and the CMB constraint is evaded; see Eq.~\eqref{turning_point_CS} or Fig.~8 of Ref.~\cite{Chang:2021afa}.


\section{GW signatures from an Early Matter-Dominated Era}
\label{sec:matter_era}

We consider a period of early matter domination (EMD) inside the usual radiation era.
We parametrize the EMD era by two parameters: 1. the temperature $T_{\rm dec}$ of the thermal plasma when the EMD ends, and 2. the duration of the EMD characterized by the number of e-folds $N_{\rm MD}$ of cosmic expansion,
\begin{align}
\exp(N_{\rm MD}) \equiv \frac{a(T_{\rm dec})}{a(T_{\rm dom})} = \left[\frac{g_*(T_{\rm dom})}{g_*(T_{\rm dec})}\right]^{1/3} \left(\frac{T_{\rm dom}}{T_{\rm dec}}\right)^{4/3},
\label{eq:nMD_tempeerature}
\end{align}
where $T_{\rm dom}$ is the radiation temperature when the EMD era starts. During a matter era, the universe expands faster than during a radiation era, inducing a \emph{double-step} and \emph{single-step} suppression of the GW spectrum from local and global strings, respectively, see Fig.~\ref{fig:spectrum_strings}. 

\begin{figure}[!ht]
\centering
\includegraphics[width=0.495\textwidth, scale=1]{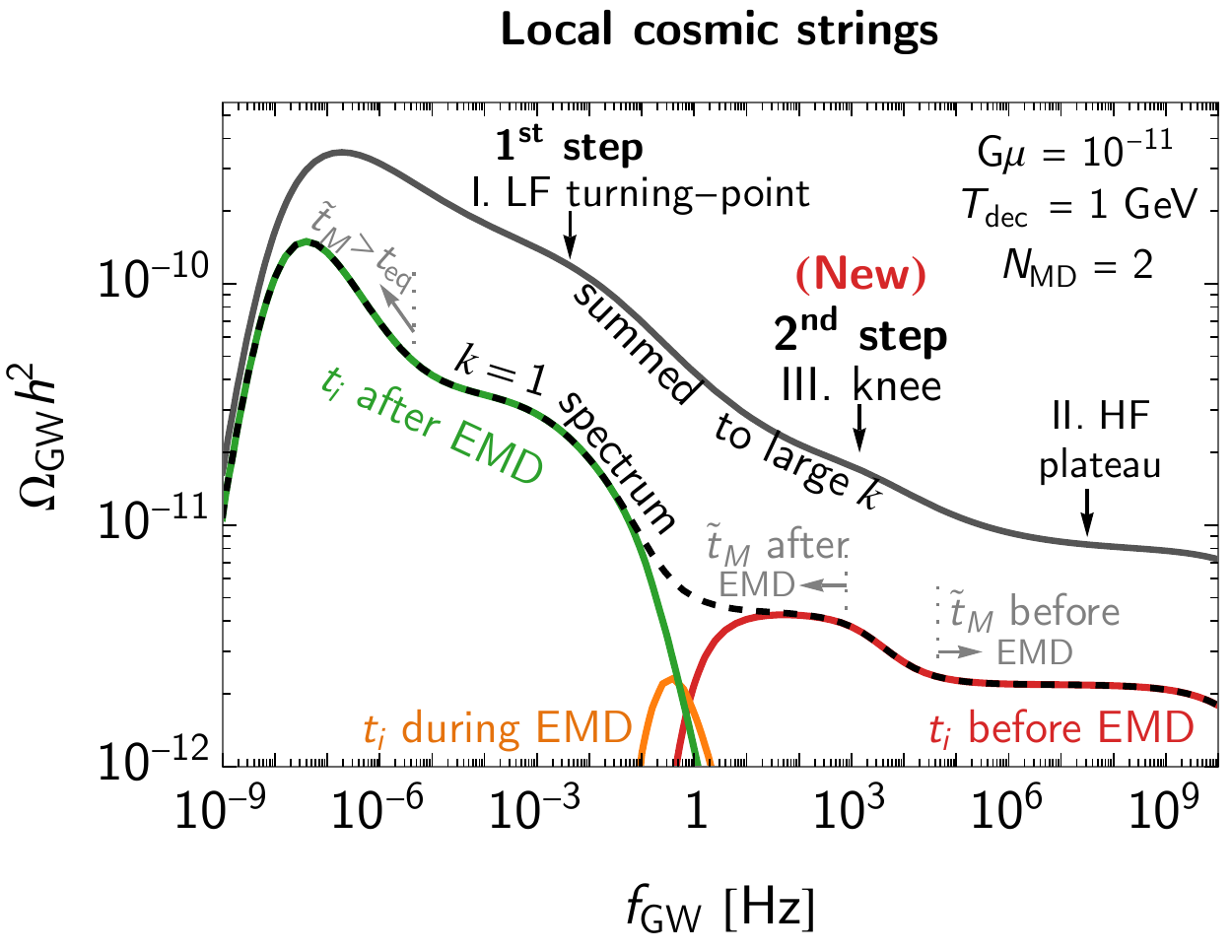}
\hfill
\includegraphics[width=0.495\textwidth, scale=1]{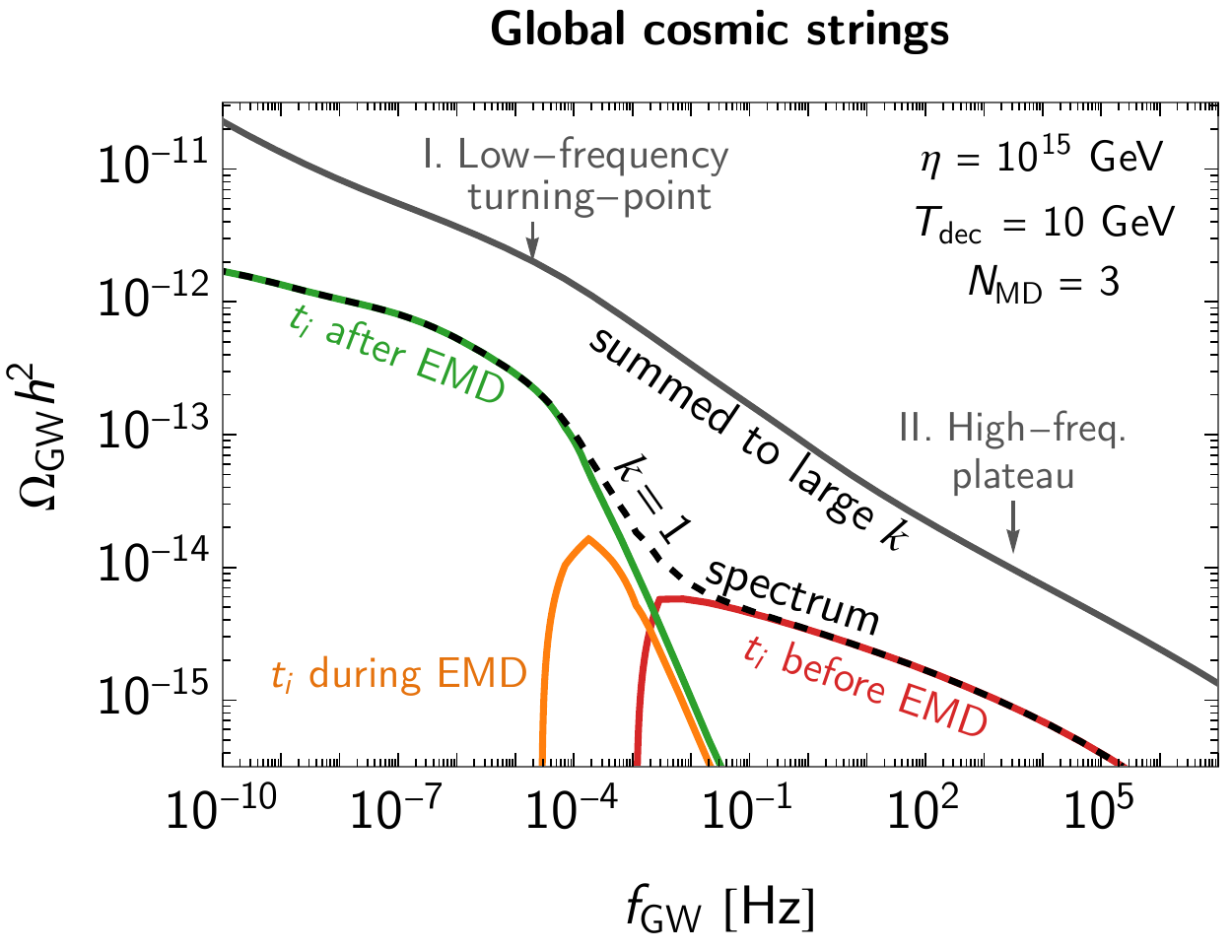}
\\[-1em]
\caption{\textit{ \small \textbf{Left:} Three features in the local-string GW spectrum induced by the EMD era. Their origins can be understood, from the $(k=1)$ spectrum, as different loop populations produced at time $t_i$ and emitting GW at time $\tilde{t}_M$ before and/or after the EMD era.  The GW spectrum from the local string shows a double-step feature. The steep slope of the 1st step is caused by loops formed during the matter era. The steep slope of the 2nd step results from loops decaying during the matter era. The knee is defined as the local maximum of the second step. \textbf{Right:} For global strings,  the GW spectrum exhibits a single step due to the loop lifetime being short.}}
\label{fig:spectral_feature}
\end{figure}

In this section, we scrutinize these \emph{step} signals and find three smoking-gun features -- shown in Fig.~\ref{fig:spectral_feature} -- which carry direct information about the EMD:
\begin{center}
    \begin{tabular}{ccc}
    {\bf Features} & {\bf EMD Info.} & {\bf Expressions} \\ \hline
    I. Low-frequency (LF) turning point & $T_{\rm dec}$ & Eq.~\eqref{turning_point_CS} \\
    II. High-frequency (HF) plateau &  $N_{\rm MD}$ &       Eq.~\eqref{eq:HF_plateau_amplitude} \\
    III. \emph{Knee} (only local strings) & $T_{\rm dec}$,  $N_{\rm MD}$ & Eqs.~\eqref{eq:knee_amplitude}--\eqref{eq:knee_frequency}.
    \end{tabular}
\end{center}
In the following subsections, we discuss their origins, derive the value of the frequency at their position, and calculate their detectability in future GW observatories. We emphasize that the present paper is the first to introduce the \emph{double-step} feature; see Sec.~\ref{sec:knee_feature} for more details. 

\subsection{Spectral index}
\label{sec:spectral_index}
\textbf{Cosmic-strings GW via a \emph{two-step} process.}-- As mentioned in Sec.~\ref{sec:GW_CS}, the cosmic-string network first produces string loops whose energy density red-shifts as non-relativistic matter, $\rho_{\rm loop} \propto a^{-3}$. Loops dominantly contribute to the GW spectrum at a time $\tilde{t}_M$. It is defined by the time when the string length has shrunk by a factor two $l(\tilde{t}_M)=l(t_i)/2$; see Eq.~\eqref{eq:tilde_t_M}.
The fraction of energy density in GW today from cosmic-string loops -- produced at time $t_i$ and emitting GW at time $\tilde{t}_M$ -- is 
\begin{align}
\Omega_{\rm GW}^{\rm CS}(f) =\left.\left(\frac{\rho_{\rm GW}}{\rho_{\rm tot}}\right)\right|_0 \simeq \frac{\rho_{{\rm loop},i}}{\rho_{{\rm tot},i}} \cdot \left(\frac{H_{i}}{H_{ 0}}\right)^2 \left(\frac{a_{i}}{a_{M}}\right)^3 \left(\frac{a_{M}}{a_{\rm 0}}\right)^4 \propto \left(\frac{H_{i}}{H_{ 0}}\right)^2 \left(\frac{a_{i}}{a_{M}}\right)^3 \left(\frac{a_{M}}{a_{\rm 0}}\right)^4,
\label{eq:simple_omegaGW}
\end{align}
where $i$, $M$, and $0$ denote the epochs of loop formation, main GW emission, and today, respectively. These three epochs are related through Eq.~\eqref{eq:tilde_t_M} and $f \simeq (4k/(\alpha t_i)) (a_M/a_0)$, cf. Eq.~\eqref{eq:emitted_frequency}. 
In Eq.~\eqref{eq:simple_omegaGW}, we have used that $\rho_{\rm loop,M} \simeq \rho_{\rm GW,M}$ (from energy conservation), 
$\rho_{\rm loop} \propto a^{-3}$, $\rho_{\rm GW} \propto a^{-4}$, and  $\rho_{{\rm loop},i} \propto \rho_{{\rm tot},i}$ (due to the scaling regime).
The GW amplitude today in Eq.~\eqref{eq:simple_omegaGW} is sensitive to the cosmic history around the time $t_i$ of loop production and the time $\tilde{t}_M$ of GW emission.
For global strings, the short loop lifetime implies that the two processes are simultaneous, and the GW emission occurs as soon as the loop is formed $\tilde{t}_M\simeq t_i$.

\textbf{Fundamental Fourier mode only.} --
We assume that the universe evolves as $\rho_{\rm tot} \propto a^{-n}$ and $a^{-m}$ around loop formation at $t_i$ and dominant GW emission at $\tilde{t}_M$ respectively.
The first Fourier mode $k=1$ of the GW spectrum in Eq.~\eqref{eq:master_eq_ready_to_use} can be expressed as the power-law 
\begin{align}
    \Omega_{\rm GW}^{\rm CS} \propto \mathcal{B}(f) f^\beta,
    \label{eq:cosmic_arhaeology_CS}
\end{align}
where $\mathcal{B}(f)$ and $\beta$ have analytical expressions \cite{Gouttenoire:2021jhk} (see also \cite{Cui:2017ufi, Cui:2018rwi, Gouttenoire:2019kij, Chang:2019mza, Chang:2021afa, Simakachorn:2022yjy}),\footnote{In particular, see \cite{Simakachorn:2022yjy} for a derivation using Eq.~\eqref{eq:simple_omegaGW}.}
\begin{center}
{\renewcommand{\arraystretch}{1.8}%
\begin{tabular}{c|c|c}
 & {local} & {global} \\ \hline
$\mathcal{B}(f)$ & $1$ & $\log^3\left[ (5.6 \cdot 10^{30}) \left(\frac{\eta}{10^{15} ~ {\rm GeV}}\right) \left( \frac{1 ~ {\rm mHz}}{f_{\rm dec}}  \right)^2 \left(\frac{f_{\rm dec}}{f}\right)^{n/(n-2)} \right]$ \\[0.5em]
$\beta$ &  $2\left[\frac{3m + n(1- m)}{n(2-m)}\right]$ & $2\left(\frac{n-4}{n-2}\right)$,
\end{tabular}}
\end{center}
where $f_{\rm dec}$ is the GW frequency related via Eq.~\eqref{turning_point_CS} to the temperature when the Universe becomes radiation-dominated again. 
The GW spectrum exhibits different characteristics depending on when the loop formation and emission occur in the history of the universe. For instance, when both events occur during radiation domination with $n=m=4$, the GW spectrum is flat in frequencies $f^0$. On the other hand, when a loop forms during matter domination with $m=3$ and emits during radiation domination with $n=4$, the resulting spectrum has a tilt of $f^{-1}$. Finally, when loop formation and GW emission happen during matter domination with $m=n=3$, the analytical prediction for the spectral tilt is $f^{-2}$. This result contradicts numerical simulations, as shown in Fig.~\ref{fig:spectrum_strings}, where $f^{-1}$ is observed. The reason for this discrepancy is that the assumption of a single time $\tilde{t}_M$ for GW emission is no longer valid in this regime.

\textbf{Effect from higher Fourier modes.} --
The discussion in the paragraph above only includes the fundamental mode $k = 1$.  Effects from higher Fourier modes have been shown in \cite{Blasi:2020wpy, Gouttenoire:2019kij, Chang:2021afa} to lead to a departure from $f^{-1}$,
\begin{equation}
\Omega_{\rm GW}^{\rm CS}  \propto \begin{cases}
f^0,\qquad \quad f \lesssim f_{\rm dec},  \\
f^{\delta-1}, \qquad f_{\rm dec} \lesssim f \lesssim k_{\rm max}f_{\rm dec},  \\
f^{-1}, \qquad  ~f \gtrsim k_{\rm max}f_{\rm dec},
\end{cases}
\end{equation}
where $k_{\rm max}$ is the maximal excited Fourier mode, which can be quite large cf. Eq.~\eqref{eq:maxmodek} in App.~\ref{app:mode_max_perturbative}, and $\delta $ is the spectral tilt of the GW emission power of a string loop, cf. Eq.~\eqref{eq:power_emisssion_GW_strings}. In this paper, we assume a cusp-dominated small-scale structure for which $\delta = 4/3$ \cite{Olmez:2010bi}. We obtain the spectral index $f^{-1/3}$. 
For long matter era, we indeed observe a spectral slope $-1/3$, see Fig.~\ref{fig:spectral_feature} and \ref{fig:local_string_slope}. For the short matter era, a more complex spectral shape emerges, with a distinctive \emph{knee}.

\begin{figure}[!ht]
\centering
\includegraphics[width=0.49\textwidth, scale=1]{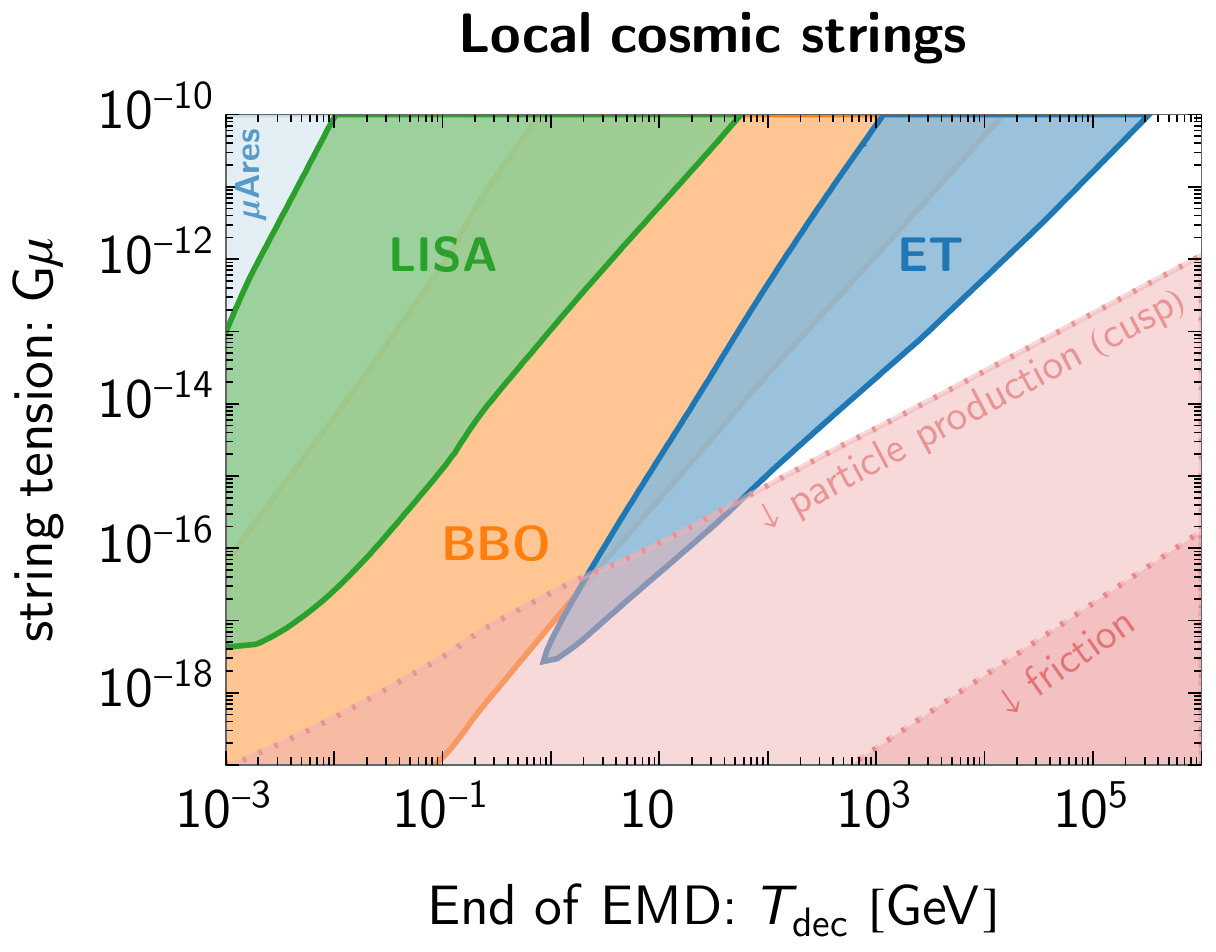}
\hfill
\includegraphics[width=0.485\textwidth, scale=1]{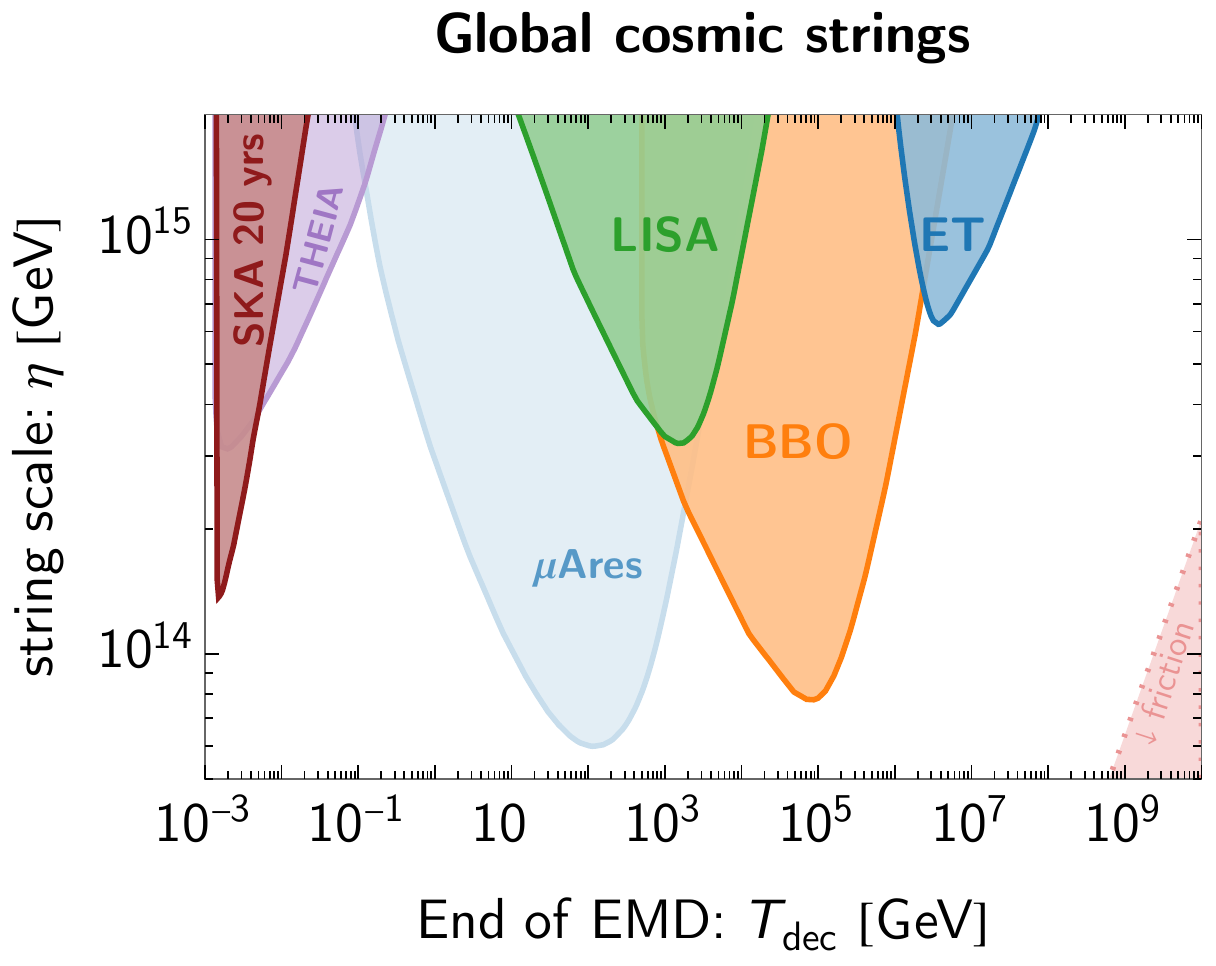}\\[-1em]
\caption{\textit{ \small Detectability of the low-frequency turning point, associated with $T_{\rm dec}$ when the EMD era ends; cf. Eq.~\eqref{turning_point_CS}. The regions correspond to the detection of ${\rm SNR} \geq 10$. The red regions correspond to the cusp and friction cut-offs in Eqs. \eqref{eq:freq_string_friction_cutoff} and \eqref{UVcutoff_f_app} that erase the LF turning point.} }
\label{fig:Gmu_vs_Tdec}
\end{figure}

\subsection{Low-frequency turning point}
The low-frequency (LF) turning point separates GWs emitted by loops formed during the radiation era, characterized by a slope $f^{0}$, from those formed during the matter era, characterized by a slope $f^{\beta}$ with $0<\beta \leq 1/3$. The frequency $f_{\rm dec}$ of the LF turning point in Eq.~\eqref{eq:master_eq_ready_to_use} is sensitive to the temperature $T_{\rm dec}$ at which the EMD era ends.
Fig.~\ref{fig:Gmu_vs_Tdec} shows the parameter space where future GW experiments can detect the LF turning point for both local and global cosmic strings. The detection criterion for this analysis is SNR $\geq 10$, obtained by comparing the GW signal with the power-law integrated sensitivity curves defined in App.~\ref{app:sensitivity_curve}. The red regions in the figure indicate where the particle production and friction cut-offs, given by Eqs. \eqref{eq:freq_string_friction_cutoff} and \eqref{UVcutoff_f_app}, respectively, lies at higher frequencies than $f_{\rm dec}$.

\subsection{High-frequency plateau}
\label{sec:hf_plateau}
The plateau at high-frequency results from GW emitted during the radiation era preceding the EMD era at a temperature $T_{M,\rm HF} >T_{\rm dom}$. $T_{M,\rm HF}$ is the temperature when loop produced at $T_{\rm HF}$ release most of their energy into GW.
From Eq.~\eqref{eq:simple_omegaGW}, GW in the HF plateau is suppressed with respect to GW at the LF turning point by
\begin{align}
    \frac{\Omega_{\rm GW}^{\rm HF-plateau}}{\Omega_{\rm GW}^{\rm dec}} &\simeq \left(\frac{H_{\rm HF}}{H_{\rm dec}}\right)^2 \left(\frac{a_{M,\rm HF}}{a_{M,\rm dec}}\right)^4 
    \left(\frac{a_{\rm HF}}{a_{M,\rm HF}}\right)^3 \left(\frac{a_{M,\rm dec}}{a_{\rm dec}}\right)^3,\nonumber\\
    &\simeq \left(\frac{a_{\rm dom}}{a_{\rm dec}}\right) \left(\frac{a_{\rm dec}}{a_{M, \rm dec}}\right) \left(\frac{a_{M,\rm HF}}{a_{\rm HF}}\right) \left(\frac{\mathcal{G}(T_{\rm HF})}{\mathcal{G}(T_{\rm dom})}\right),\nonumber\\
    &\simeq \exp(-N_{\rm MD}) \left[\frac{\mathcal{G}(T_{\rm dec}) \mathcal{G}(T_{M, \rm HF}) \mathcal{G}^3(T_{ \rm HF})}{\mathcal{G}(T_{M,\rm dec}) \mathcal{G}^{4}(T_{\rm dom})}\right]^{1/4},
    \label{eq:HF_plateau_amplitude}
\end{align}
where the second step uses $\rho_{\rm MD} \propto a^{-3}$ and $\rho_{\rm RD} \propto a^{-4}\mathcal{G}(T)$ with $\mathcal{G}(T)$ in Eq.~\eqref{eq:Delta_R}, and the third step uses  $a(t) \propto \mathcal{G}^{1/4}(T) t^{1/2}$ (from Friedmann's equation) and the loop lifetime in Eq.~\eqref{eq:tilde_t_M}. $T_{M,\rm dec}$ is the temperature when loop produced at $T_{\rm dec}$ release most of their energy into GW. The function $\mathcal{G}(T)$  varies from $\simeq 0.4$ at high temperatures to 1 at low temperatures. The last bracket reduces to $\mathcal{O}(1)$ and 1 for local and global strings respectively. The suppression of the HF plateau encoded in Eq.~\eqref{eq:HF_plateau_amplitude} is visible in Fig.~\ref{fig:spectrum_strings}.
In App.~\ref{app:spectrum_without_dof_evo}, we also show the cosmic-string GW spectrum when the number of degrees of freedom $g_*$ and $g_{*s}$ are fixed, i.e. taking $\mathcal{G}(T)\equiv 1$.
For the global string, the HF plateau $f^0$ has a distinct HF turning point at $f = f_{\rm dec}$ below which the slope turns to $f^{1/3}$.  In App.~\ref{app:hf_turning_point_global}, we provide an analytic formula for the HF turning point of global strings and argue that it is difficult to detect, e.g., see Fig.~\ref{fig:hf_turning_point_global}.  
For local string, the spectral slope below $f<f_{\rm dec}$ is not exactly $-1/3$ due to the `\emph{knee}' feature, which we now discuss below.

\begin{figure}[!ht]
\centering
{\bf Local cosmic strings}\\[0.25em]
\centering
\includegraphics[width=0.495\textwidth, scale=1]{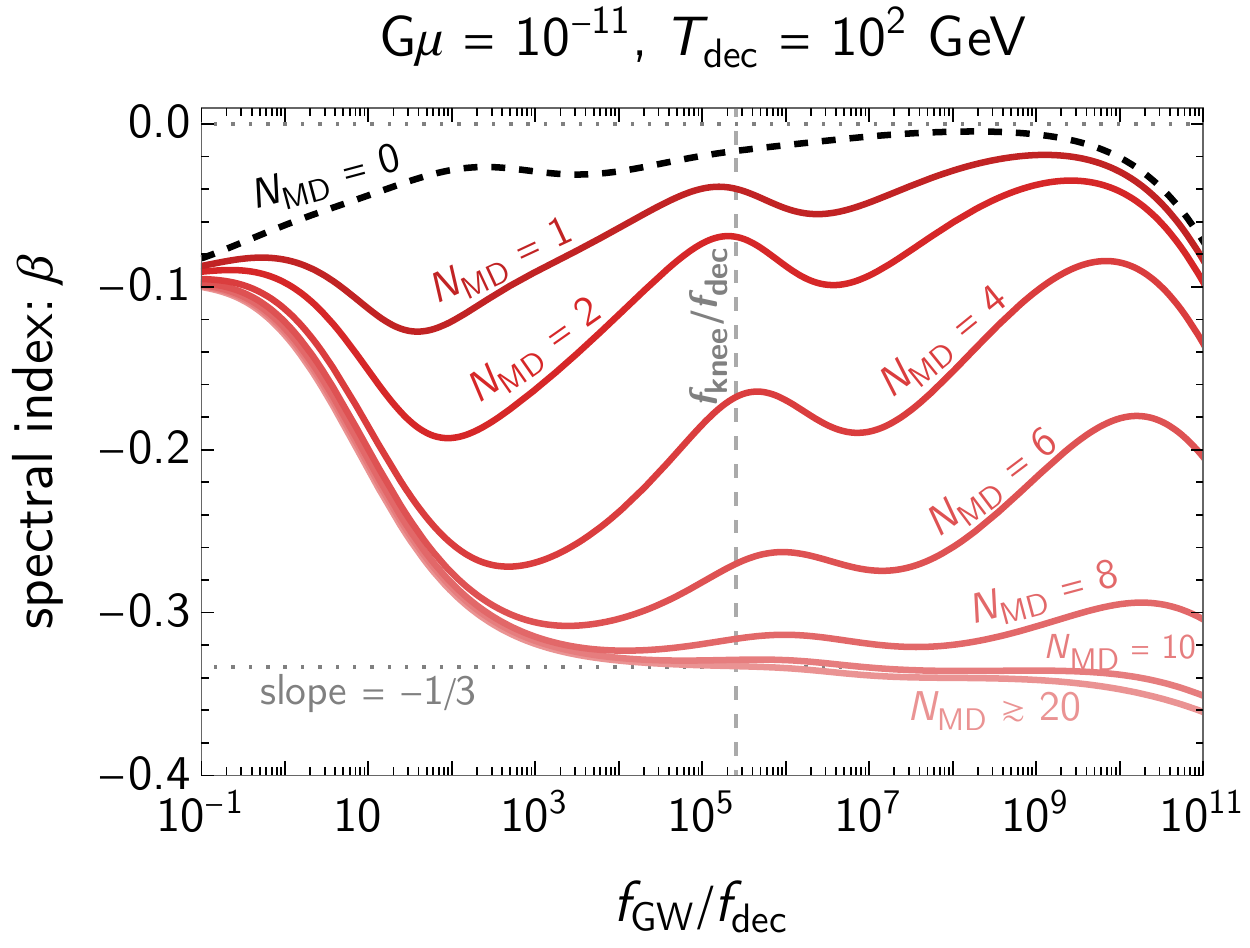}
\hfill
\includegraphics[width=0.495\textwidth, scale=1]{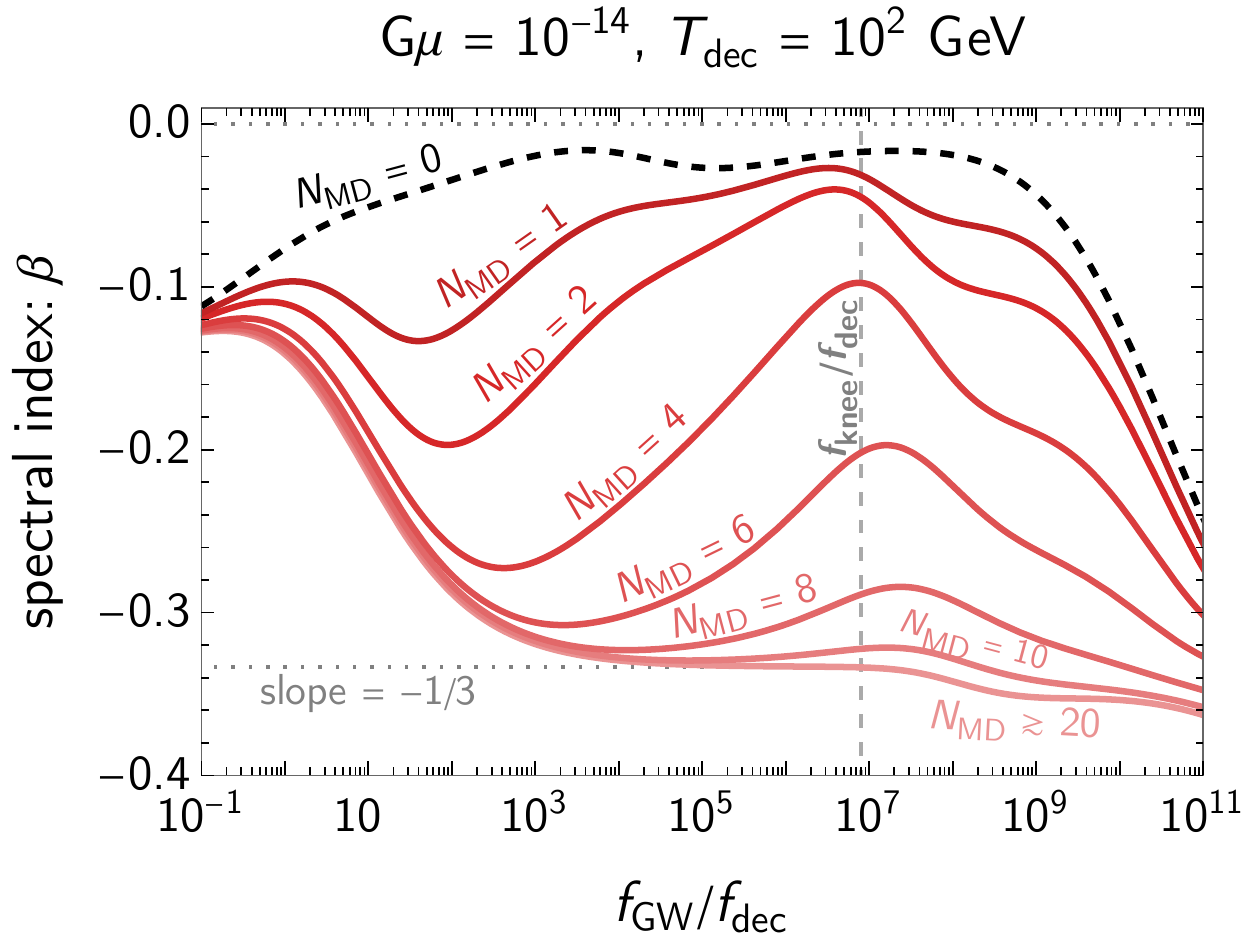}
\\[-1em]
\caption{\textit{ \small Numerical result of the spectral slope (i.e., $\beta \equiv d \log \Omega_{\rm GW}/d \log f$) of the local-string GW  experiencing different duration $N_{\rm MD}$ of the EMD era. 
The vertical dashed lines show the position of the \emph{knee} feature, specified by Eq.~\eqref{eq:knee_frequency}.
Only for $N_{\rm MD} \gtrsim 10$ (cf. App.~\ref{app:knee_feature_visibility}), the \emph{knee} feature submerges below the $-1/3$-slope tail from the LF turning point. 
}}
\label{fig:local_string_slope}
\end{figure}

\begin{figure}[!ht]
\centering
{\bf Local cosmic strings}\\[0.25em]
\includegraphics[width=0.49\textwidth, scale=1]{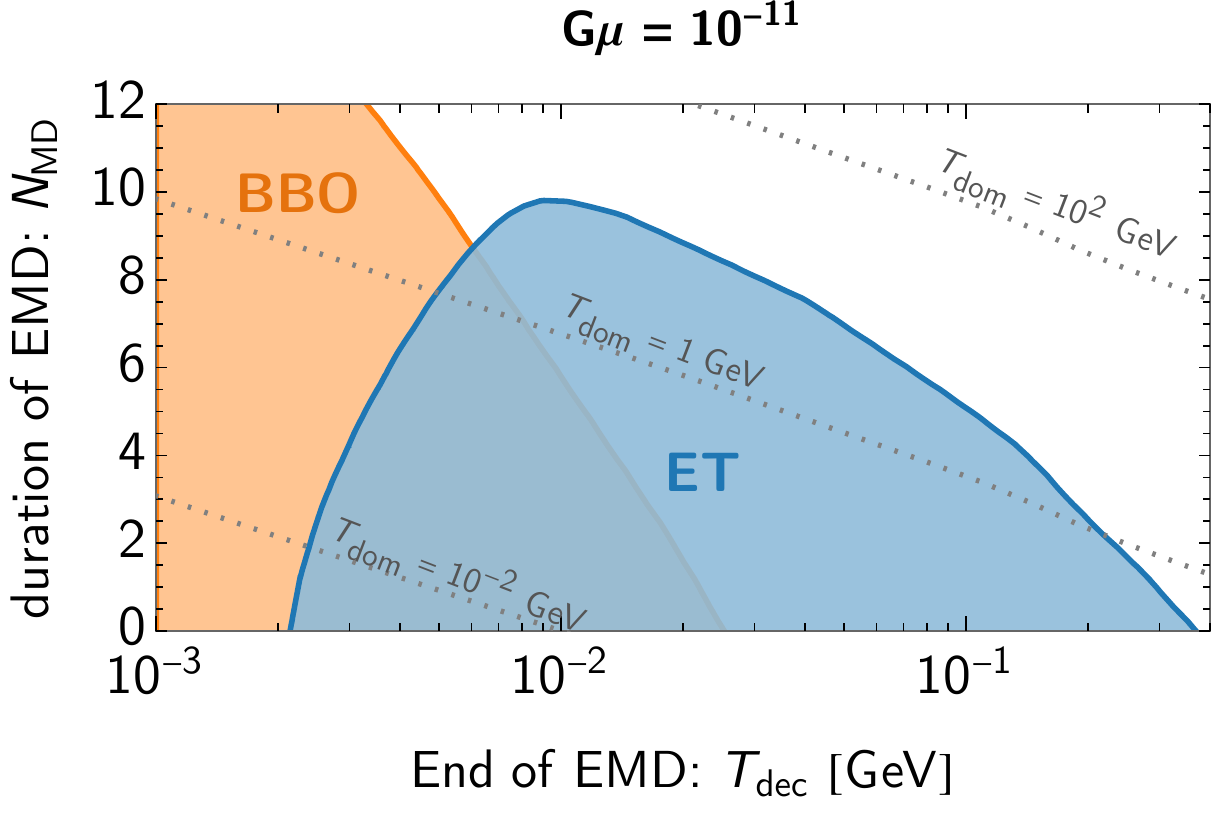}
\hfill
\includegraphics[width=0.49\textwidth, scale=1]{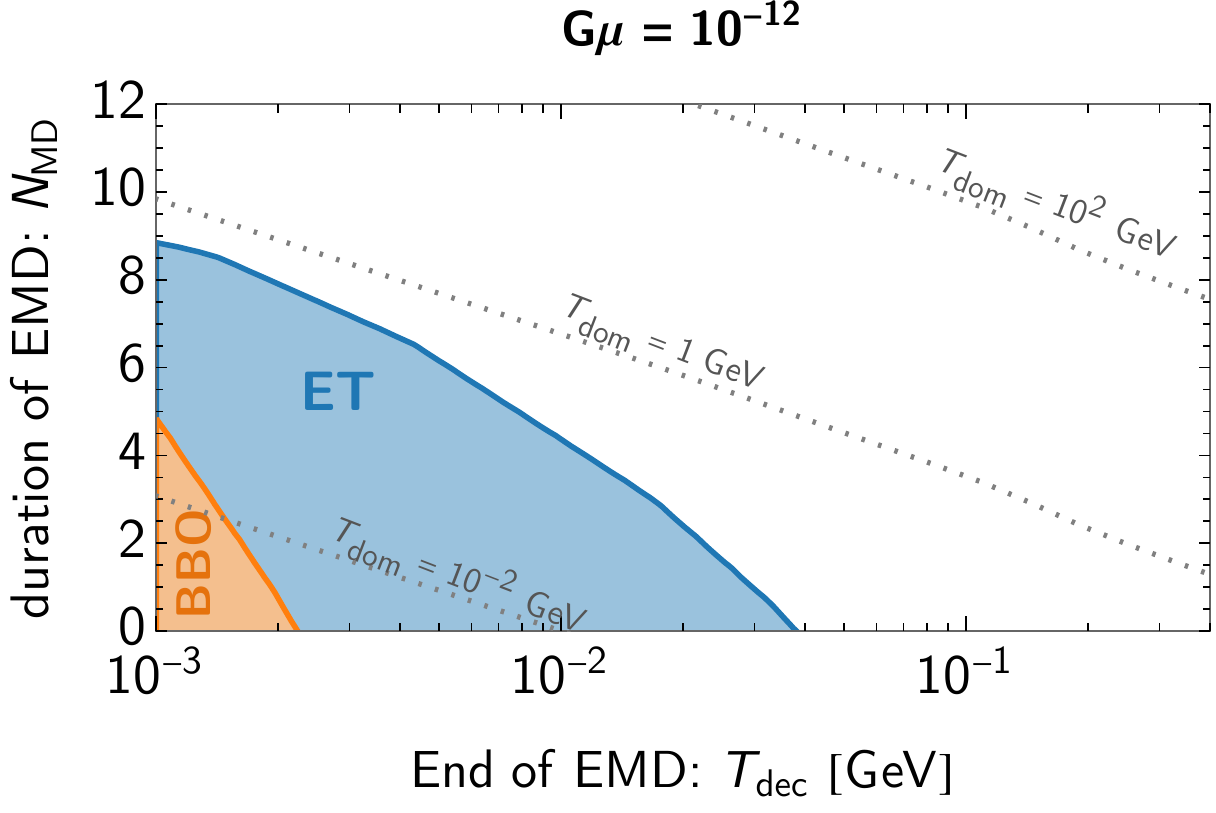}\\[-1em]
\caption{\textit{ \small Detectability of the \emph{knee} feature at future GW observatories. Its observation provides information about the energy scale and duration of the EMD. We cut the parameter space to $N_{\rm MD} < 12$ because the \emph{knee} visibility is lost for larger $N_{\rm MD}$; see App.~\ref{app:knee_feature_visibility} for more details. } }
\label{fig:knee_feature}
\end{figure}

\subsection{The knee feature}
\label{sec:knee_feature}
In the case of local strings, the GW spectrum has an additional feature due to loops living longer than the duration of the EMD. In Fig.~\ref{fig:spectral_feature}, we show for the first time that the EMD imprints not only one but two steps in the GW spectrum. They originate from the production of GWs occurring in two steps: first, loops are formed at a time $t_i$, and second, they decay at a later time $\tilde{t}_M$. The first step suppression in the GW spectrum is attributed to the formation of loops during the EMD, while the higher-frequency step suppression results from the decay of loops during the EMD. The intermediate region of the GW spectrum is sourced by string loops that formed prior to the onset of the EMD era and emitted most of their energy into GWs after the EMD era ended. Fig.~\ref{fig:local_string_slope} shows the numerical results of the spectral slopes for different $G\mu$ and EMD durations.
We call the \emph{knee} the local maximum of the second step, between the LF turning point and the HF plateau; see the vertical dashed line. The \emph{knee} feature can also be seen clearly in Figs.~\ref{fig:spectrum_strings} and \ref{fig:spectral_feature}. Such features do not show up in the GW spectrum from global strings because the short lifetime of loops merges the two steps into one.

Although the \emph{knee} feature is the smoking-gun signature of the EMD
era, the same underlying physics can also leave an imprint on the local-string GW spectrum, assuming
the standard $\Lambda$CDM history \cite{Blanco-Pillado:2017oxo}, shown in Fig.~\ref{fig:spectrum_strings}.
The spectral peak -- located around $f_{\rm GW} \simeq 150 ~ {\rm nHz} ~ (50 \cdot 10^{-11})/(\Gamma G\mu)$ \cite{Gouttenoire:2019kij} -- has its UV slope from loops with $t_i < t_{\rm eq}< \tilde{t}_M$ and its IR cut-off due to the matter-$\Lambda$ transition. Nonetheless, no knee feature is present for the standard cosmological history due to the lack of a double-step spectrum, unlike the EMD case.

We now analytically estimate the position of the \emph{knee} feature and calculate its detectability by future GW observatories.
The spectral slope reaches its maximum -- the tip of the \emph{knee} -- when the time $\tilde{t}_M$ of dominant GW emission occurs at the very end of the EMD $\tilde{t}_M =t_{\rm dec}$; see Figs.~\ref{fig:spectral_feature} and \ref{fig:knee_feature}.
Using Eq.~\eqref{eq:simple_omegaGW} with $a_{M, \rm knee} = a_{\rm dec}$, we obtain an analytic estimation for the GW amplitude at the \emph{knee},
\begin{align}
    \frac{\Omega_{\rm GW}^{\rm knee}}{\Omega_{\rm GW}^{\rm dec}} &\simeq \left(\frac{H_{\rm knee}}{H_{\rm dec}}\right)^2 \left(\frac{a_{\rm dec}}{a_{M,\rm dec}}\right)^4 
    \left(\frac{a_{\rm knee}}{a_{\rm dec}}\right)^3 \left(\frac{a_{M,\rm dec}}{a_{\rm dec}}\right)^3,\nonumber\\
    &\simeq \left(\frac{a_{\rm dom}}{a_{\rm knee}}\right) \left(\frac{a_{\rm dec}}{a_{M, \rm dec}}\right) \left(\frac{\mathcal{G}(T_{\rm knee})}{\mathcal{G}(T_{\rm dom})}\right),\nonumber\\
    &\simeq \exp\left(-
    3N_{\rm MD}/4\right) \left[\frac{\mathcal{G}(T_{\rm dec}) \mathcal{G}^3(T_{\rm knee})}{ \mathcal{G}(T_{M, \rm dec}) \mathcal{G}^3(T_{\rm dom})}\right]^{1/4},
    \label{eq:knee_amplitude}
\end{align}
where the second step uses $\rho_{\rm MD} \propto a^{-3}$ and $\rho_{\rm RD} \propto a^{-4}\mathcal{G}(T)$ with $\mathcal{G}(T)$ in Eq.~\eqref{eq:Delta_R}, and the third step uses  $a_{\rm RD}(t) \propto \mathcal{G}^{1/4}(T) t^{1/2}$, $a_{\rm MD}(t) \propto t^{2/3}$, and the loop lifetime in Eq.~\eqref{eq:tilde_t_M}.
Applying Eqs.~\eqref{eq:emitted_frequency} and \eqref{eq:tilde_t_M}, we arrive at the \emph{knee} frequency,
\begin{align}
    f_{\rm knee} &\simeq f_{\rm dec} \left(\frac{t_{\rm dec}}{t_{\rm knee}}\right) \left(\frac{a_{\rm dec}}{a_{M, \rm dec}}\right) \simeq f_{\rm dec} \left(\frac{\alpha}{2 \Gamma G \mu}\right)^{1/2} \left[\frac{\mathcal{G}(T_{\rm dec})}{\mathcal{G}(T_{M, \rm dec})}\right]^{1/4},\nonumber\\
    &\simeq 607 \, {\rm Hz} \, \left(\frac{50 \cdot 10^{-11}}{\Gamma G \mu}\right) \left(\frac{T_{\rm dec}}{\rm GeV}\right) \left[\frac{g_*^2(T_{\rm dec})}{g_*(T_0) g_*(T_{M, \rm dec})}\right]^{\frac{1}{4}} \left[\frac{g_{*s}(T_{M, \rm dec})}{g_{*s}(T_{\rm dec})}\right]^{\frac{1}{3}}
    \label{eq:knee_frequency}
\end{align}
where we have used $f_{\rm dec}(T_{\rm dec})$ in Eq.~\eqref{turning_point_CS} multiplied by a numerical factor of 33.5)\footnote{In Eq.~\eqref{turning_point_CS}, $f_{\rm dec}(T_{\rm dec})$ has already taken into account of the network (VOS) evolution and the smoothening effect from the high Fourier modes which shift the apparent LF turning point lower than the one expected from the analytic estimate from the $k=1$ spectrum by a factor of 1/33.5; see Eq.~28 of Ref.~\cite{Gouttenoire:2019kij}. However, our estimation of the knee frequency in Eq.~ \eqref{eq:knee_frequency} is based on the $k=1$ spectrum; hence we will use the LF turning point of the $k=1$ spectrum.} 
We can see that Eq.~\eqref{eq:knee_frequency} describes well the numerical results in Fig.~\ref{fig:spectrum_strings}. In App.~\ref{app:spectrum_without_dof_evo}, we show GW spectrum at fix $g_*$ and $g_{*s}$ in order to make effects from the \emph{knee} more visible.

The \emph{knee} feature described by Eqs.~\eqref{eq:knee_amplitude} and \eqref{eq:knee_frequency} is sensitive to both the duration $N_{\rm MD}$ of the EMD era and its end temperature $T_{\rm dec}$, unlike the low-frequency turning point feature which only depends on the latter. By detecting the \emph{knee} feature, future GW observatories could reconstruct the full EMD era, as shown in Fig.~\ref{fig:knee_feature}. Even if the \emph{knee} feature lies outside the detectability region, Fig.~\ref{fig:local_string_slope} indicates that observing a portion of the GW spectrum at frequencies $f>f_{\rm dec}$ would enable determination of the EMD duration $N_{\rm MD}$. While an analytical expression for the spectral index in Fig.~\ref{fig:local_string_slope} is challenging to derive due to the effects of higher Fourier modes and GW emission beyond the dominant epoch $\tilde{t}_M$, it remains a topic for future studies.

The presence of a second step in the GW spectrum already appeared in previous numerical calculations, e.g., Fig.~12 of Ref.~\cite{Gouttenoire:2019kij} and Fig.~9 of Ref.~\cite{Gouttenoire:2021jhk},\footnote{Non-trivial spectral tilts can also arise in the presence of multiple epochs of EMD \cite{Borah:2022vsu}.} but its origin and detectability are studied for the first time in the present paper. This knee feature can potentially serve as a distinguishing characteristic of the EMD signature from the effects of particle production cutoffs \cite{Auclair:2019jip}, which also have a $f^{-1/3}$ slope or supercooling phase transition effects \cite{Ferrer:2023uwz} which has peak plus a plateau feature in the GW spectrum. A short intermediate inflation \cite{Guedes:2018afo, Gouttenoire:2021jhk, Cui:2019kkd} could also lead to similar behaviors, we leave its study for future work.

\section{Primordial Black Holes Domination Era}
\label{sec:PBH_era}

After they form in the early Universe, PBHs may be abundant enough to dominate the total energy density before they evaporate through Hawking radiation. Because their mass is relatively constant before Hawking evaporation becomes sizeable, they behave like a component of cold matter in the early Universe. In this section, we explore the possibility that the early matter domination era studied in the previous section, and its effect on the GW spectrum emitted by cosmic strings, are due to the existence of such PBHs. 

\subsection{PBH Formation}
PBHs form as a consequence of density fluctuations present in the early universe. When the density in a region exceeds a certain threshold, the gravitational forces become strong enough to overcome the Hubble expansion and pressure \cite{Carr:1975qj}, leading to gravitational collapse and the formation of a black hole.  
 These fluctuations can arise from various mechanisms, e.g. \cite{Carr:1993aq,Ivanov:1994pa,Hawking:1982ga,Moss:1994iq,Moss:1994pi,Crawford:1982yz,Gross:2021qgx,Baker:2021sno,Kawana:2021tde,Sato:1981bf,Maeda:1981gw,Sato:1981gv,Kodama:1981gu,Kodama:1982sf,Garriga:2015fdk,Deng:2016vzb,Deng:2017uwc,Dolgov:1992pu,Dolgov:2008wu,Kitajima:2020kig,Kasai:2022vhq,Martin:2019nuw,Martin:2020fgl,Vachaspati:2017hjw,Ferrer:2018uiu,Hawking:1987bn,Polnarev:1988dh,Fort:1993zb,
Garriga:1993gj,Caldwell:1995fu,MacGibbon:1997pu,Jenkins:2020ctp,Blanco-Pillado:2021klh}.
Some scenarios produce PBH distributions that peak at a particular mass, while others result in extended distributions related to the Fourier spectrum of the primordial fluctuations and to the equation of state of the Universe at the time they collapsed ~\cite{Carr:1975qj, Harada:2013epa,Harada:2016mhb,Carr:2017jsz, Cheek:2022mmy, Heurtier:2022rhf}.  Our study avoids making assumptions about the specific mechanism of overdensity formation and simplifies the analysis by considering a nearly-monochromatic density distribution.
\begin{equation}
\frac{dn_{\rm PBH}}{dM_{\rm PBH}}\propto \delta(M-M_{\rm PBH})\,.
\end{equation}
We introduce the parameter $\gamma$, which represents the fraction of mass within the Hubble horizon that collapses into a PBH. We can then establish a relationship between the mass $M_{\rm PBH}$ of the PBH and the temperature $T_f$ at which it forms
\begin{equation}
\label{eq:MPBH_at_formation}
M_{\rm PBH}  = \gamma \rho\frac{4\pi}{3} \left(\frac{1}{H(T_f)}\right)^3 \simeq 10^9~{\rm g}~ \left( \frac{\gamma}{0.2}\right) \left( \frac{100}{g_*(T_f)}\right)^{1/2}\left( \frac{1.4 \times 10^{11}~\rm GeV}{T_f}\right)^{2},
\end{equation}
where $g_*(T_f)$ is the number of relativistic degrees of freedom in the plasma at temperature $T_f$.
For PBH collapse occurring from super-horizon fluctuations entering the horizon during RD, we have $\gamma \simeq c_s^3 \simeq 0.2$ where $c_s = 1/\sqrt{3}$ is the speed of sound in a relativistic plasma. 
The energy fraction of PBHs at formation is defined as
\begin{equation}
\label{eq:beta_parameter}
\beta(M_{\rm PBH}) \equiv \frac{\rho_{\rm PBH}(T_f)}{\rho_{\rm tot}(T_f)},
\end{equation} 
where $\rho_{\rm tot}(T_f)$ denotes the total energy density in the Universe when the temperature of the SM plasma equals $T_f$.

\subsection{PBH Evaporation}
The presence of a Schwarzschild horizon implies that PBHs emit a distribution of particles that can be well approximated by a thermal distribution with temperature \cite{Hawking:1974rv, Hawking:1974sw}
\begin{equation}
T_{\rm PBH} = \frac{1}{8\pi G M_{\rm PBH}} \simeq 1.06~{\rm GeV} ~\left(\frac{10^{13}~\rm g}{M_{\rm PBH}}\right).
\end{equation} 
The corresponding production rate for particle $j$ is
\begin{equation}
\label{eq:production_rate_specie}
\frac{dN_j}{dt dE} = \frac{g_j}{2\pi} \frac{\Gamma_j(E,\,M_{\rm PBH})}{e^{E/T_{\rm PBH}}-(- 1)^{s_j}},
\end{equation}
where $g_j$ is the number of internal degrees of freedom, $s_j$ is its spin, and $\Gamma_j(E,\,M_{\rm PBH})$ is the greybody factor~\cite{Page:1976df,Page:1976ki}.  As a result of Hawking evaporation, the PBH mass decreases at a rate
\begin{equation}
\label{eq:mass_decrease_rate}
\frac{dM_{\rm PBH}}{dt} = - \sum_j \int_0^\infty dE\, E \frac{dN_j}{dtdE} = -\varepsilon(M_{\rm PBH})\frac{M_{\rm pl}^4}{M_{\rm PBH}^2},
\end{equation}
where $M_{\rm pl} \simeq 2.44 \times 10^{18}~\rm GeV$ and the function $\varepsilon$  is a function that encodes the details of the Hawking emission which depends on the particle physics spectrum considered and the Hawking temperature of the black hole (for a thorough description including the PBH spin, see Refs.~\cite{Cheek:2021cfe, Auffinger:2022khh}). 
 In this study, we assume the emitted particles to belong to the SM only,\footnote{In the case of global strings, the massless scalar boson should, in principle, be added to the discussion. However, it would only affect the results regarding the evaporation temperature by $\mathcal O(1\%)$, and can thus be safely ignored here.}, and since we focus on PBHs with masses $M_{\rm PBH}\lesssim 10^9~\mathrm{g}$ corresponding to Hawking temperatures $T_{\rm PBH}\gtrsim 10^4~\mathrm{GeV}$, all the SM degrees of freedom can be assumed to be relativistic, giving the constant evaporation rate $\varepsilon\simeq 4.4\times 10^{-3}$.

Upon integrating Eq.~\eqref{eq:mass_decrease_rate} over time, one straightforwardly obtains the lifetime of a PBH with mass $M_{\rm PBH}$ at formation:
\begin{equation}\label{eq:PBHlifetime}
\tau(M_{\rm PBH})  = \frac{1}{3\varepsilon}\frac{M_{\rm PBH}^3}{M_{\rm pl}^4} \simeq 0.41~ {\rm s}~\left ( \frac{M_{\rm PBH}}{10^{9}~{\rm g} }\right)^3.
\end{equation}

\begin{figure}[!ht]
\centering
{\bf Gravitational waves from local cosmic strings}\\[0.25em]
\centering
\includegraphics[width=0.75\textwidth, scale=1]{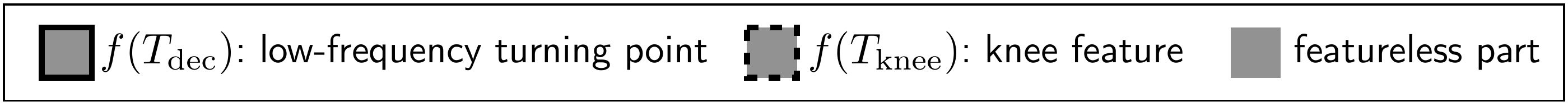}\\[0.1em]
\includegraphics[width=\textwidth, scale=1]{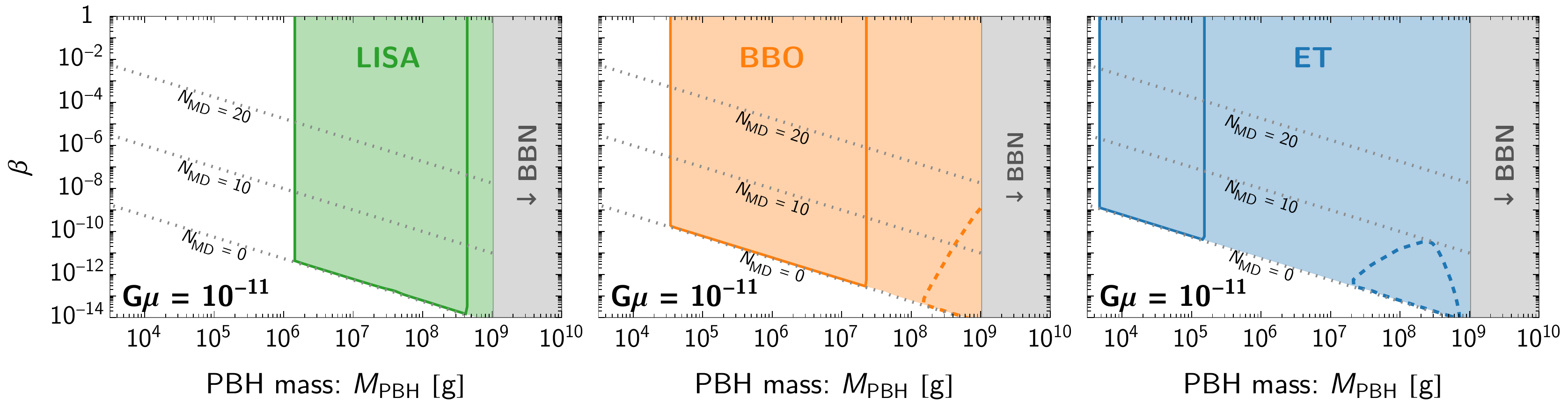}\\
\includegraphics[width=\textwidth, scale=1]{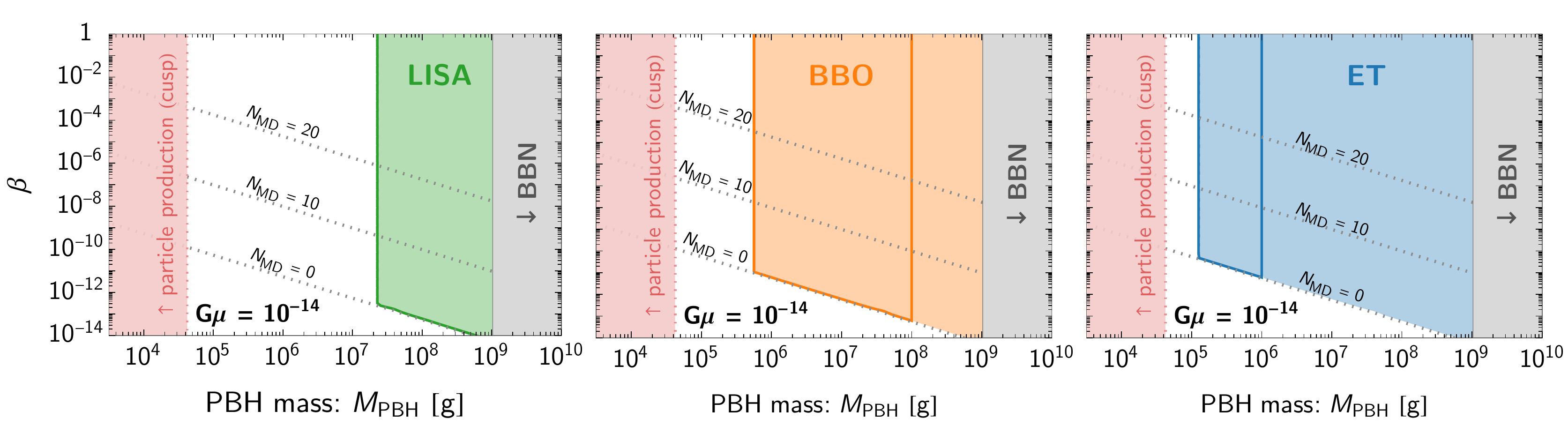}\\[-1em]
\caption{\label{fig:local_pbh_plane}\textit{ \small Detectability of the double-step signature from the PBH domination. The regions bounded by the solid and dashed lines correspond to a detectable LF turning point and the knee feature. The unbounded colored region is where the ``featureless" part is detectable (information about the duration of PBH domination can be retrieved by measuring the spectral slope, shown in Fig.~\ref{fig:local_string_slope}). }}
\end{figure}

\begin{figure}[!ht]
\centering
{\bf Gravitational waves from global cosmic strings}\\
\includegraphics[width=0.55\textwidth, scale=1]{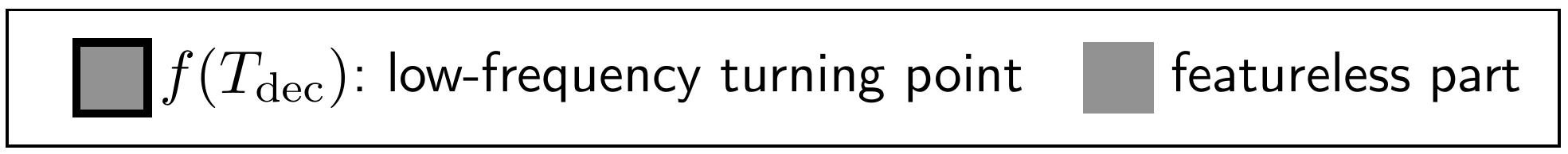}\\[0.1em]
\includegraphics[width=\textwidth, scale=1]{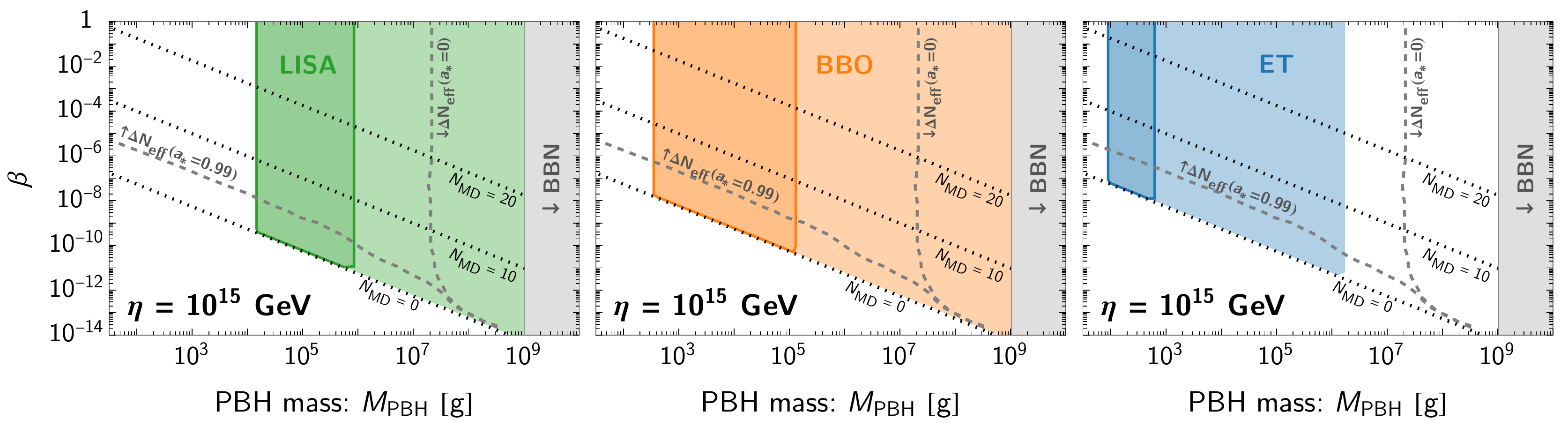}\\
{Detectable LF turning point}\\
\includegraphics[width=\textwidth, scale=1]{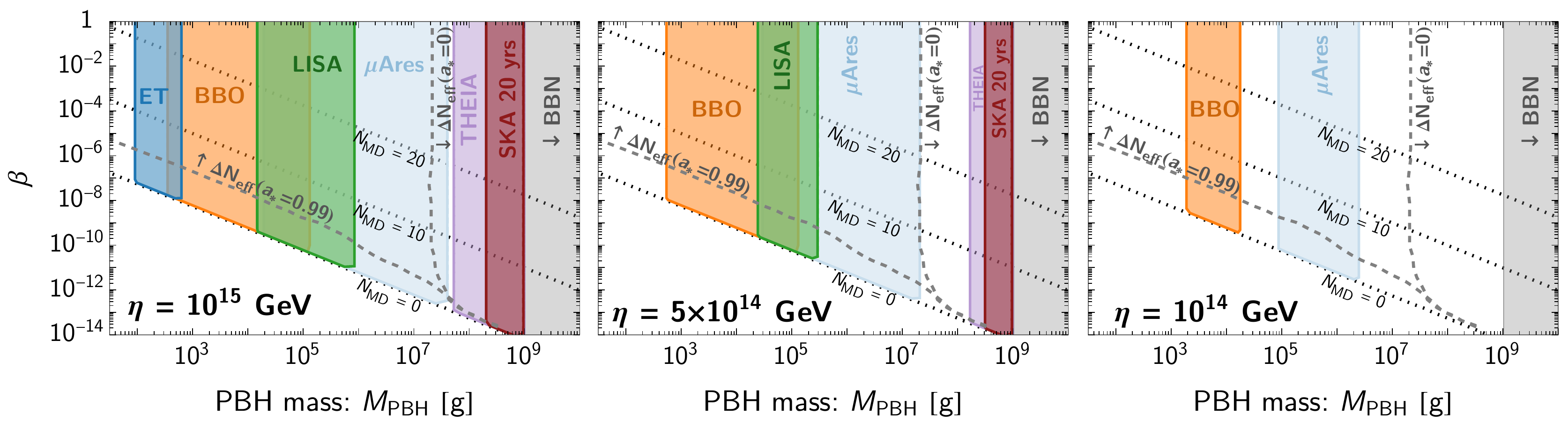}\\[-1em]
\caption{\textit{\small Top panel: Detectability of the single-step feature from PBH domination: LF turning point (bounded by solid line) and the featureless part (unbounded). Bottom panel: Detectable LF turning point associated with the end of PBH domination in the GW background from global cosmic strings. This is the solely detectable signature on the global-string GW spectrum as the HF turning point locates at the ultra-high frequency and the spectral index is always $-1/3$. Dash lines indicate the $\Delta N_{\rm eff}$ possibly observable by CMB-HD due to the massless goldstone of the global-string $U(1)$ symmetry produced by the evaporation PBH of spin parameter $a_*$.}}
\label{fig:global_pbh_plane}
\end{figure}

\begin{figure}[!ht]
\centering
\includegraphics[width=0.49\textwidth, scale=1]{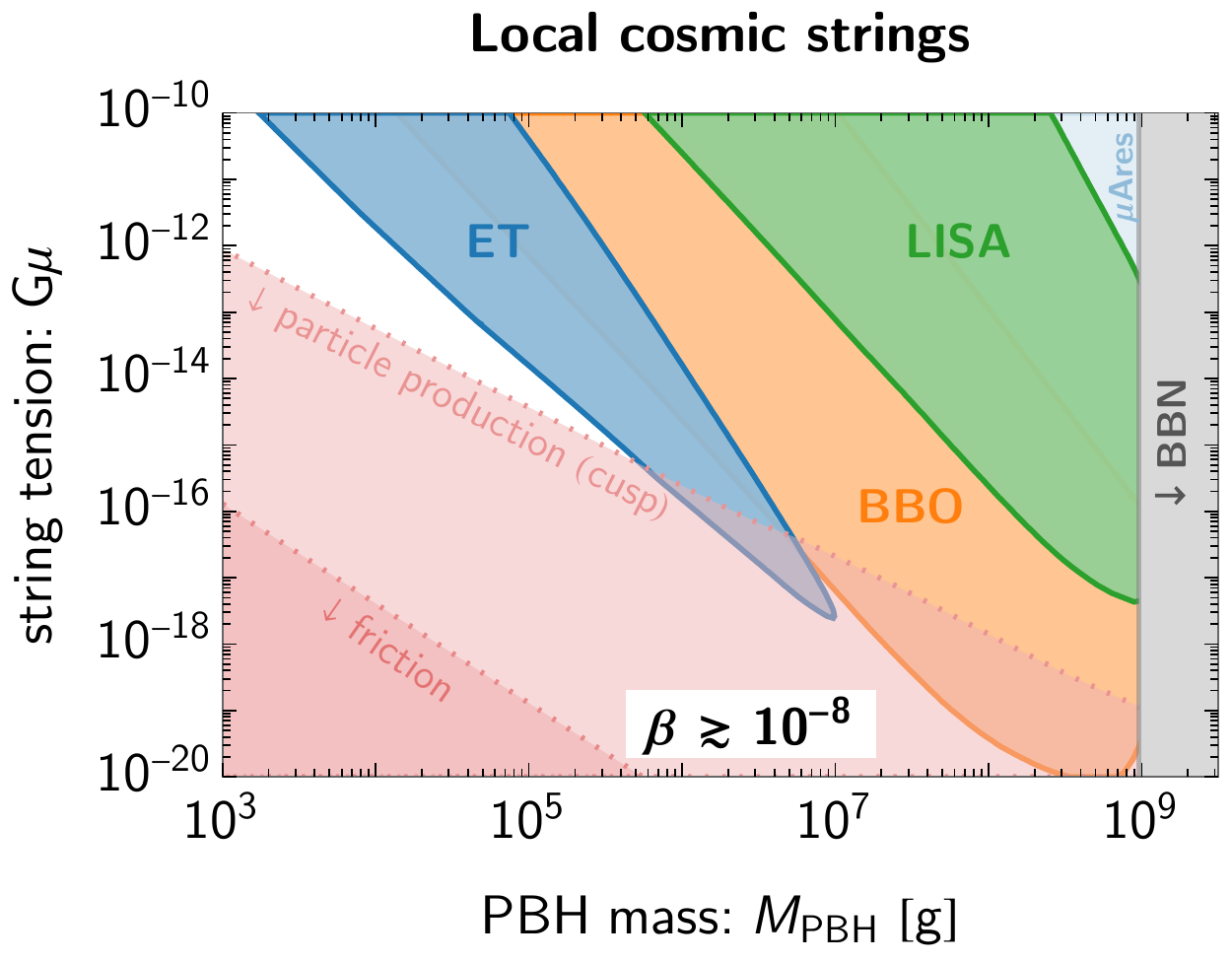}
\hfill
\includegraphics[width=0.485\textwidth, scale=1]{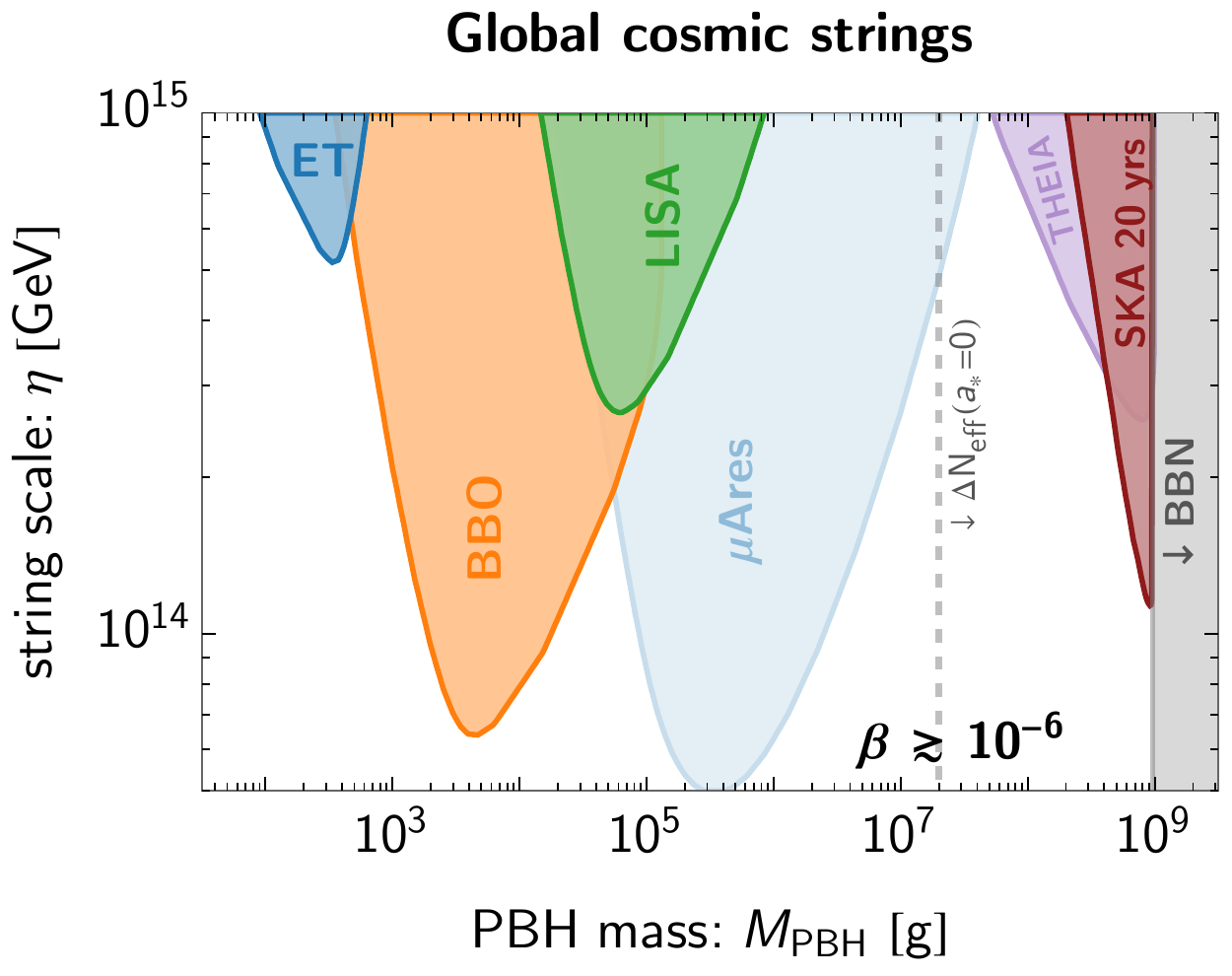}\\[-1em]
\caption{\textit{ \small  Detectability of the LF turning point due to the PBH domination with varying $G\mu$ (local) and $\eta$ (global). For global strings, if PBHs have a spin parameter $a_\star = 0.99$, almost the whole parameter space may be probed by CMB-HD; see the previous figure.} }
\label{fig:just_check}
\end{figure}

\subsection{PBH-Dominated Era}

PBHs smaller than $10^{15}~\rm g$ (150 times the Great Pyramid of Giza) are supposed to have already evaporated by now due to Hawking radiation. PBHs smaller than $10^{9}~\rm g$ (one-tenth of the Eiffel tower) have even evaporated before the onset of BBN, leaving not many observables left to probe their existence.
However, if their energy fraction $\beta$ at formation introduced in  Eq.~\eqref{eq:beta_parameter} is sufficiently large, PBHs may end up dominating the energy density of the Universe before they evaporate, hence leading to an EMD era. 
Using that PBHs redshift like matter, we obtain that PBHs domination starts when the temperature of the SM plasma drops below
\begin{eqnarray}
T_{\rm dom} \,=\, T_f \beta = 14 \,{\rm GeV}\left(\frac{\beta}{10^{-10}}\right)\! \left(\frac{100}{g_{*}(T_{\rm f})} \right)^{1/4}\!\left( \frac{\gamma}{0.2} \right)^{1/2}\!\left(\frac{10^9~\rm g}{M_{\rm PBH}} \right)^{1/2},
\end{eqnarray}
where we have used Eq.~\eqref{eq:MPBH_at_formation}. After they dominate the total energy density, PBHs reheat the Universe when they evaporate. The reheating temperature $T_{\rm dec}$ immediately after evaporation is the temperature of a radiation-dominated universe whose age is equal to the PBH lifetime in Eq.~\eqref{eq:PBHlifetime}:
\begin{equation}
\label{eq:Tdec_exp}
T_{\rm dec} \simeq \frac{2.4 ~\rm MeV}{g_*^{1/4}}\left ( \frac{10^{9}~{\rm g} }{M_{\rm PBH}}\right)^{3/2}.
\end{equation}
The condition $T_{\rm dom} > T_{\rm dec}$ needed to have an early period of PBH domination is equivalent to demanding the energy fraction $\beta$ to be larger than the critical value
\begin{equation}
\beta~ >~ \beta_c \equiv \frac{T_{\rm dec}}{T_f} \simeq 5.5 \times 10^{-15}\left( \frac{0.2}{\gamma} \right)^{1/2} \left( \frac{g_{*,s}(T_{f})}{g_*(T_{\rm dec})} \right)^{1/4} \left( \frac{10^{9}~\rm g}{M_{\rm PBH}} \right).
\end{equation}
The total number of $e$-folds of the corresponding early matter-domination era starting at temperature $T_{\rm dom}$ and ending at temperature $T_{\rm dec}$ can then be simply obtained as 
\begin{equation}
N_{\rm MD}  = 10 + \log\left[ \left( \frac{\gamma}{0.2} \right)^{1/2} \left( \frac{g_{*,s}(T_{\rm dec})}{g_*(T_{f})} \right)^{1/4} \left( \frac{M_{\rm PBH}}{10^{9}~\rm g} \right)\left( \frac{\beta(t_f)}{1.2\times 10^{-10}} \right) \right].
\end{equation}
As discussed in Sec.~\ref{sec:matter_era}, an early matter-domination era would induce a change of slope in the GW spectrum emitted from a pre-existing cosmic string network.
In Fig.~\ref{fig:local_pbh_plane}, \ref{fig:global_pbh_plane}, and \ref{fig:just_check}, we show the detection prospects of the imprint of a period of PBH domination by future interferometers.

We consider the PBH-dominated era to be detectable if it leads to a suppression of the GW spectrum larger than 10\% as compared to standard cosmology. Using Eq.~\eqref{eq:knee_amplitude}, one can translate this condition into a threshold on the number of $e$-folds of PBH domination to be $N_{\rm MD} > 0.14$ or $T_{\rm dom}/T_{\rm dec} \gtrsim 1.11$.

As shown in Fig.~\ref{fig:local_pbh_plane}, measuring GWs emitted by local-strings with $G\mu = 10^{-11}$ can allow probing the existence of PBHs with masses in the range $[10^{6}, 10^9]$ g at LISA and $[5 \times 10^{3}, 10^9]$ g at ET. The LF turning-point signature (solid lines) in each detector probes the PBH mass towards the small mass range, while the knee feature (dashed lines) is associated with a larger PBH mass. Interestingly, joint efforts between the different collaborations could allow us to accurately pin down the PBH parameters, e.g., for $G\mu \simeq 10^{-11}$, a PBH of mass $\mathcal{O}(10^{8})$ g, and $\beta \sim 10^{-10}$,  LISA could observe the LF turning point while ET observes the knee. In the unbounded colored region labeled ``featureless part", we show the detectability of the GW spectrum outside the LF turning point and the knee. As shown in Fig.~\ref{fig:local_string_slope}, the non-trivial spectral slope in those regions also carries information about the duration of the PBH domination era.

The ability to probe the PBH parameter space with global-string GWs is shown in Fig.~\ref{fig:global_pbh_plane}. The top panel provides the observational bounds on the LF turning point and the \emph{featureless} part, while the bottom panel gives more details on the former. 
From the top panel of Fig.~\ref{fig:global_pbh_plane}, PBHs with masses $[10^4,10^9]$ g and $[10^2, 10^6]$ g can be probed by LISA and ET, respectively, with global strings of $\eta = 10^{15} \, {\rm GeV}$. Instead, the bottom panel shows other GW observatories sensitive to LF turning points at frequencies lower than LISA. In contrast to GWs emitted by local strings, whose frequency corresponding to the BBN scale is $10^{-6}$ Hz, the GWs emitted by global strings can have a LF turning point at smaller frequencies; see Eq.~\eqref{turning_point_CS}.
In this figure, we also show for comparison bounds on PBHs that could be set in the future by CMB-HD, using the fact that massless goldstone bosons may be produced copiously from the evaporation of PBHs and contribute to $\Delta N_{\rm eff}$. As one can see from the figure, in the case of rotating PBHs with spin parameter $a_\star=0.99$, GW searches and CMB observations may observe complementary smoking-gun signatures of the existence of a PBH-dominated era, whereas, for Schwarzschild PBHs, GW searches reveal to be much more competitive at low PBH masses.

Fig.~\ref{fig:just_check} shows the string-scale ($G\mu$ and $\eta$) dependence of the detectability of PBH parameters for large values of $\beta$, above which all the shown parameter spaces are guaranteed to have $N_{\rm MD} > 0$; see Figs.~\ref{fig:local_pbh_plane} and  \ref{fig:global_pbh_plane}. Note that one loses sensitivity relatively fast in the case of global strings for a decreasing $\eta$, as $\Omega_{\rm GW} \propto \eta^{4}$. Local-string GWs depend on $(G\mu)^{1/2}$. Therefore, even for local strings with a scale as small as $G\mu \sim 10^{-18}$, PBHs can be searched for efficiently by future experiments. Nonetheless, the particle production cutoff could erode the signature from EMD; see Eq.~\eqref{UVcutoff_f_app}.

\subsection{Effects from PBH spins} 
\label{sec:pbh_spin}
In the early Universe, PBHs may have acquired angular momentum from merging events \cite{Hooper:2020evu}, matter accretion \cite{DeLuca:2019buf}, evaporation \cite{Calza:2021czr}, primordial inhomogeneities \cite{DeLuca:2019buf}, collapse during matter domination \cite{Harada:2016mhb,Harada:2017fjm}, or specific mechanisms of PBH formation, such as scalar fragmentation \cite{Cotner:2019ykd,Flores:2021tmc}, collapse of domain walls \cite{Eroshenko:2021sez} or cosmic strings \cite{Jenkins:2020ctp,Blanco-Pillado:2021klh}. 
In that case, the PBHs can be characterized, on top of their mass, by the spin parameter $a_\star=J M_{\rm pl}^2/M_{\rm PBH}^2$ where $J$ is the angular momentum of the black hole. The  PBH spin impacts a variety of particle-physics phenomena: e.g., the production of dark matter through evaporation~\cite{Cheek:2021odj, Cheek:2021cfe, Hooper:2019gtx, Masina:2020xhk,Morrison:2018xla, Auffinger:2020afu, Khlopov:2004tn, Allahverdi:2017sks, Lennon:2017tqq, Gondolo:2020uqv, Baldes:2020nuv, Bernal:2020bjf, Bernal:2020ili, Masina:2021zpu, Kitabayashi:2021hox, Bernal:2021bbv,Bernal:2020kse,Bernal:2022oha},  the amount of dark radiation produced through evaporation that could contribute to $\Delta N_{\rm eff}$ in future CMB measurements~\cite{Cheek:2022dbx, Hooper:2019gtx, Hooper:2020evu, Masina:2020xhk, Arbey:2021ysg, Masina:2021zpu, Bhaumik:2022zdd,Cheek:2022dbx}, or even the spectrum of GW induced at second order in perturbation theory \cite{Bhaumik:2022zdd}. In this section, we explore the effect of the PBH spin on the PBH domination era and its signature in the cosmic-string GW spectrum. 

\begin{figure}[h!]
\centering
\includegraphics[width=0.495\textwidth, scale=1]{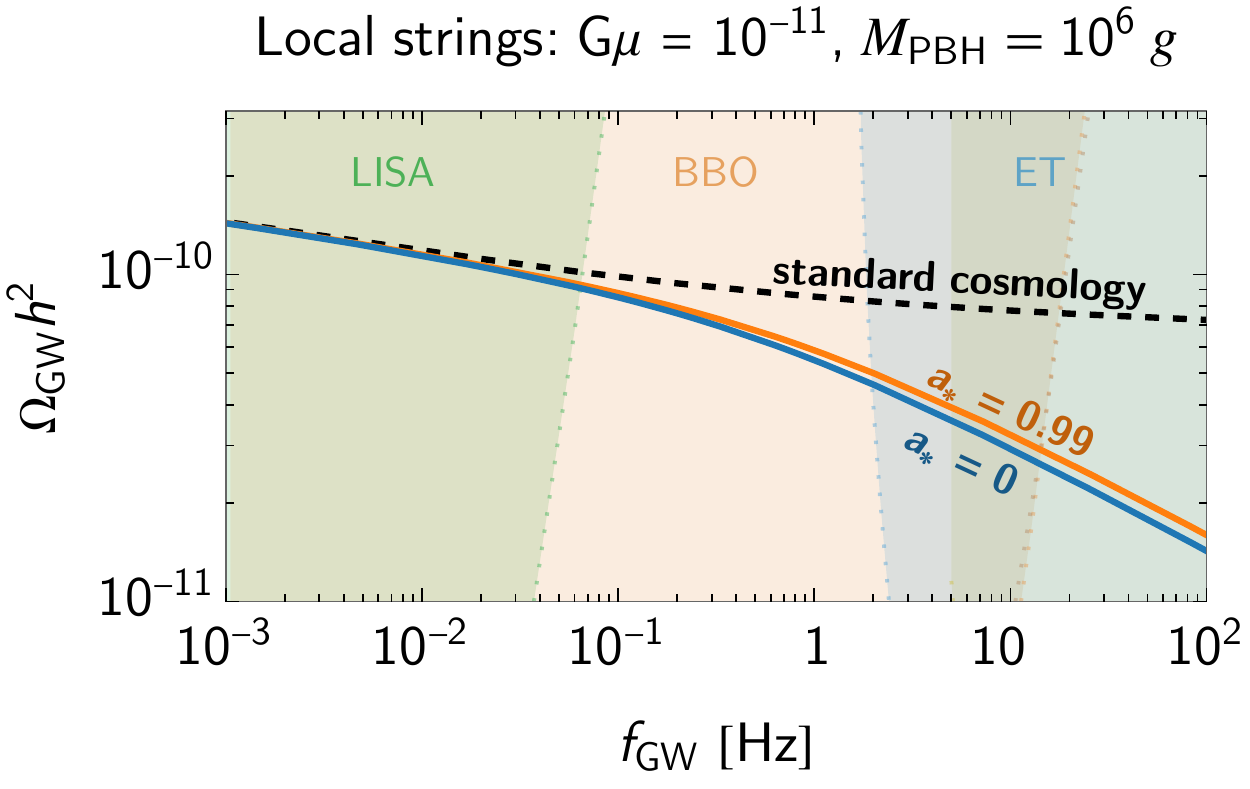}
\hfill
\includegraphics[width=0.495\textwidth, scale=1]{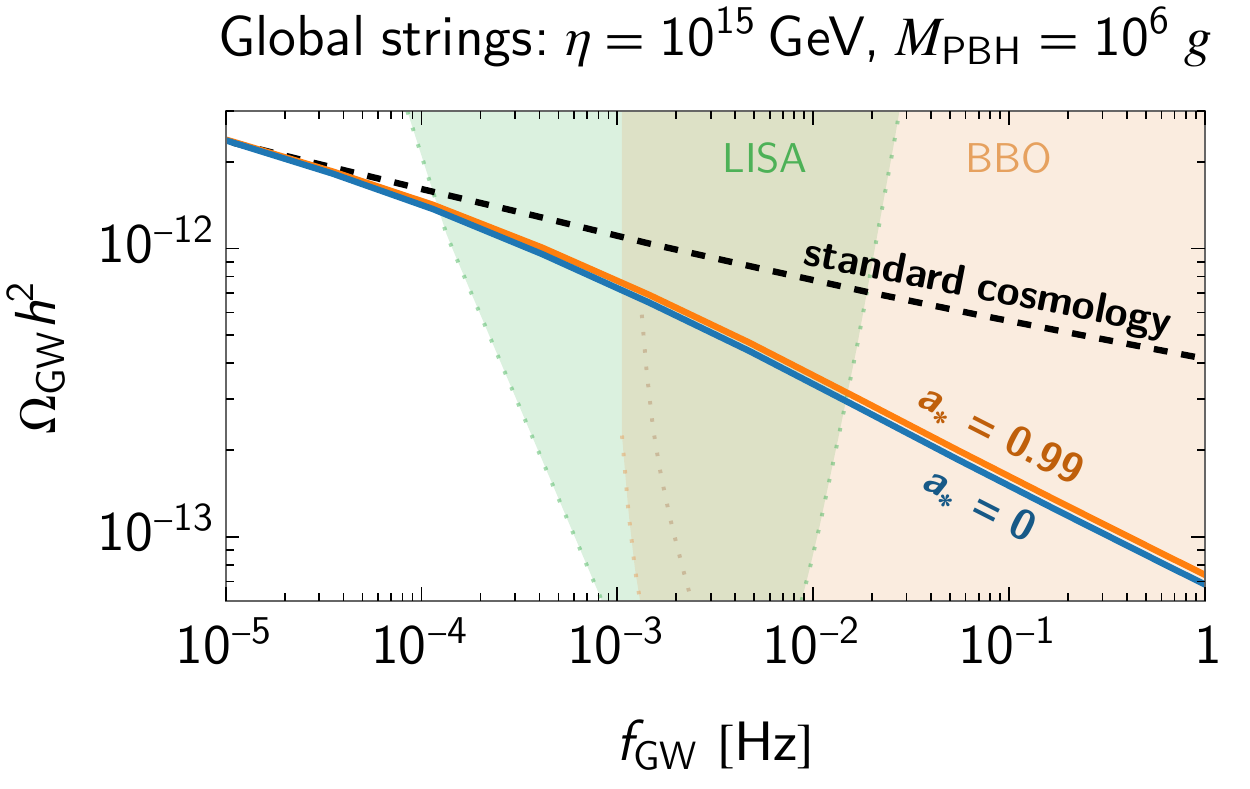}\\[-1em]
\caption{\textit{ \small Cosmic-string GW spectra experiencing PBH domination era with the monochromatic PBH mass spectrum $M_{\rm PBH} = 10^6 \, {\rm g}$ and  $\beta = 10^{-4}$. We vary the PBH spin parameter $a_*$ which affects slightly the lifetime of PBH: $T_{\rm dec} \simeq T_{\rm dec,s} \mathcal{F}^{-1/2}(a_*, M_{\rm PBH})$ for the end of Schwarzschild PBH domination $T_{\rm dec,s}$ and the function $\mathcal{F}$ is taken from Ref.~\cite{Bhaumik:2022zdd}.}}
\label{fig:spin}
\end{figure}

The spin of PBHs accelerates the dynamics of Hawking evaporation and shortens their lifetime $\Delta t_{\rm PBH}^{\rm Kerr}$ from the expectation $\Delta t_{\rm PBH}^{\rm Sch}$ assuming the Schwarzschild PBH, 
\begin{align}
    \Delta t_{\rm PBH}^{\rm Kerr}/\Delta t_{\rm PBH}^{\rm Sch} = \mathcal{F}(a_\star, M_{\rm PBH}),
\end{align}
where $\mathcal{F}(a_\star, M_{\rm PBH})$ is a numerical function which range from $\mathcal{O}(0.4)$ for $a_\star \simeq 0.999$ to $1$ for $a_\star = 0$ and taken from Table  1 of Ref.~\cite{Bhaumik:2022zdd}.
While the start of PBH domination remains unchanged by spin (fixed by the energy fraction at formation $\beta$, $\gamma$, $M_{\rm PBH}$), the spinning PBHs evaporate at a temperature higher than the Schwarzschild ones, using $\Delta t_{\rm PBH} = H_{\rm PBH}^{-1}$,
\begin{align}
    T_{\rm evap}^{\rm Kerr} / T_{\rm evap}^{\rm Sch} &= \mathcal{F}^{-1/2}(a_\star, M_{\rm PBH})\left[{g_{*}(T_{\rm evap}^{\rm Sch}) }/{g_{*}(T_{\rm evap}^{\rm Kerr})}\right]^{1/4}.
\end{align} 

The evaporation temperature of maximally-spinning PBHs ($a_\star = 0.999 \to \mathcal{F} \simeq 0.4$) is $\sim 1.6$ times higher than for static PBH, assuming for simplicity $g_{*s}(T_{\rm evap}^{\rm Kerr}) = g_{*s}(T_{\rm evap}^{\rm Sch})$. 
Concerning the signature in cosmic-string GW, Eq.~\eqref{turning_point_CS}, the turning point in the case of the maximally-spinning PBHs sits at a frequency $\sim 1.6$ times higher than in the Schwarzschild case.
In Fig.~\ref{fig:spin}, our results show that effects from spin turn out to be relatively minor. 
Furthermore, effects from the spin are degenerate with the PBH mass, and they are equivalent to decrease $M_{\rm PBH}$ by a factor $1.4$, see Eq.~\eqref{eq:Tdec_exp}.

\section{Conclusion}
\label{sec:conclusion}

The cosmological evolution of the universe below a temperature of 5 MeV is well-established and supported by numerous observational probes, including the nuclear abundances predicted by the Big Bang Nucleosynthesis (BBN) epoch, Cosmic Microwave Background (CMB), large-scale structure measurements and supernovae observations. This evolution is characterized by successive epochs of radiation domination, matter domination, and dark energy domination.
However, to explore the conditions prevailing in the pre-BBN universe with temperatures above 5 MeV, gravitational waves (GW) are one of the limited observables available. Cosmic strings are one candidate for such sources. They are expected in any high-scale theories of particle physics involving local or global U(1) symmetry breaking ranging from neutrino mass generation, leptogenesis, dark matter, flavor, or Grand Unified Theories in Beyond the Standard Model contexts \cite{Dunsky:2021tih}.
 An important particularity of cosmic strings is their scaling behavior. Their energy density scales with the scale factor precisely as the dominant energy density of the universe, i.e., $a^{-4}$ during radiation and $a^{-3}$ during matter. Consequently, the spectral slope of the GW energy density is sensitive to the equation of states of the universe at early times.

This study builds upon the research direction ``GW archaeology with cosmic strings", initiated in Refs.~\cite{Cui:2018rwi, Cui:2017ufi, Gouttenoire:2019kij, Gouttenoire:2019rtn, Blasi:2020mfx}. The objective is to use GW produced by cosmic strings to infer the energy content of the pre-BBN universe.
The present work focuses on the GW signatures of an early matter-dominated era.  The previous studies assumed that the spectral slope of the GW spectrum emitted by local cosmic strings was independent of the duration $N_{\rm MD}$ of the matter era. An important result of the present work is to reveal the contrary. Figs.~\ref{fig:spectrum_strings} and \ref{fig:local_string_slope} show that any measurement of the spectral slope would give information on the duration of the matter era. Until now, only the temperature at the end of the matter era was believed to be measurable. The present study shows that measuring the temperature at which the matter era starts is also possible.
As illustrated in Fig.~\ref{fig:spectral_feature}, the impact of the matter era leads to two-step features in the GW spectrum from local strings instead of one, as previously thought. This feature arises due to a GW emission occurring in two steps: first, a string loop is formed at time $t_i$, and second, the loop converts its energy into GW at the end of its lifetime at $\tilde{t}_M \gg t_i$. The first step in the GW spectrum is due to the impact of the matter era on the scale factor $a(t_i)$, while the second step is due to the impact of the matter era on the scale factor $a(\tilde{t}_M)$. We analytically compute the position of the second step, which we call the \emph{knee}, see Eqs.~\eqref{eq:knee_amplitude}-\eqref{eq:knee_frequency}. 
 
 Assuming the standard cosmological history, the cosmic strings with\footnote{Bounded from above by the pulsar-timing-array observations \cite{NANOGrav:2023hvm, Servant:2023mwt}.} a tension $G\mu \lesssim 10^{-11}$ (local) and an energy scale $\eta \lesssim 10^{15} ~{\rm GeV}$ (global) would be observed by LISA, ET, and other future-planned experiments. Our analysis provides a way to extract information about the EMD era, by probing any deviation from the standard-cosmology prediction. As a potential example, we show that the existence of PBHs -- if they dominate the energy density of the early universe -- can be constrained by LISA and ET for  PBH masses between $[10^6,10^9]$ g and $[5\times 10^3, 10^9]$ g, respectively, for the local strings with $G\mu = 10^{-11}$. 
Similarly in the case of global strings, LISA and ET can probe PBH masses between $[10^4,10^9]$ g and $[10^2,10^6]$ g, respectively, if  the symmetry breaking scale is at $\eta = 10^{15}~\mathrm{GeV}$.
The detectability and the ability to constrain PBHs become weaker as $G\mu$ and $\eta$ decrease because the amplitude of the SGWB is smaller, as shown in Fig.~\ref{fig:just_check}.
Finally, we considered in Sec.~\ref{sec:pbh_spin} the possibility that PBHs are Kerr black holes and therefore evaporate at a slightly earlier time than Schwarzschild black holes.

Current studies \cite{Inomata:2020lmk,Papanikolaou:2020qtd,Domenech:2020ssp,Domenech:2021wkk,Bhaumik:2022pil,Bhaumik:2022zdd, Papanikolaou:2021uhe, Papanikolaou:2022chm, Papanikolaou:2022hkg} have demonstrated that scalar-induced GW resulting from adiabatic and isocurvature perturbations can produce distinctive resonant peaks and double-peaks, which can be used to probe the formation and decay time of PBHs in the early universe. These present additional opportunities to test scenarios such as PBH domination and evaporation. The spectral shapes of those GW differ from those found in the context of GW emitted by cosmic strings studied in this work. 

Recent pulsar-timing-array data has revealed a common red-noise signal, which may be interpreted as a stochastic GW background in the frequency range $10^{-9}\, {\rm Hz}  \lesssim f \lesssim 10^{-7} \, {\rm Hz}$ \cite{NANOGrav:2020bcs, Chen:2021rqp, Goncharov:2021oub, Antoniadis:2022pcn, NANOGrav:2023gor, Antoniadis:2023rey, Reardon:2023gzh, Xu:2023wog, NANOGrav:2023hvm, Antoniadis:2023zhi}. This signal could potentially be attributed to a cosmic-string network with a tension parameter of $G\mu \in [10^{-11},10^{-10}]$ \cite{Ellis:2020ena,Blasi:2020mfx, NANOGrav:2023hvm}.
The early matter era, such as the one induced by PBHs domination, cannot distort the GW spectrum at these low frequencies, as it would require modifying the equation of state of the universe after BBN, which is at odd with concordance cosmology.
However, if the signal in the pulsar-timing-array data is caused by cosmic strings, then the high-frequency part of the SGWB spectrum will be detected by future gravitational-wave experiments.
Observing this high-frequency component would allow for constraining the presence of any early matter-dominated era occurring below $T\lesssim 10^5~\rm GeV$ and for probing the presence of PBHs population down to $10^4~\rm g$.

 There are some worth-mentioning caveats in our analysis. First, we rely on the detection criterion that requires ${\rm SNR} = 10$ by comparing our signal to the power-law integrated sensitivity curves. However, this criterion might not be enough when we confront real data. Because the detection claim of a SGWB requires a signal reconstruction over other detectors' noises at an acceptable confidence level, as shown for LISA in \cite{Caprini:2019pxz, Baghi:2023ast}. For a spectrum with several features like the \emph{knee}, the detectable signal might have to lie above the detectors' noise level in order to read the SGWB spectrum accurately and to extract precise information about the EMD era. Moreover, the optimal SNR depends on the knowledge about the noises, and one would have to perform a global fit \cite{Littenberg:2023xpl} on the combined noises and the SGWB signal.
The second caveat is that our work assumes cosmic strings and PBHs do not interact during the PBH-dominated era. However, PBHs can affect the cosmic string network through mechanisms such as chopping off long strings, modifying the network-scaling behavior, suppressing loop production, or leading to necklace-like or net-like structures \cite{Siemens:2000ty, Blanco-Pillado:2007kxw, Kibble:2015twa}. These processes could result in additional GW contributions \cite{Vilenkin:2018zol} beyond our conservative estimates. We leave the study of the string-PBH network for future work.

If the features found in this paper are observed in the GW spectrum from cosmic strings, additional observations will be needed to distinguish between a PBH-dominated era and other forms of early matter domination. It is worth noting that PBHs generate other signals that could be explored in the future. In Fig.~\ref{fig:global_pbh_plane}, we show possibilities for detecting a non-zero
component of dark radiation. Other indicators of the presence of evaporating PBHs in the early universe, such as second-order GWs, dark matter searches, baryogenesis, and structure formation, provide various ways to independently verify the existence of an early PBH-domination era (see e.g. Ref.~\cite{Auffinger:2022khh} for a recent review). Such detection channels would not only provide valuable independent confirmations of our results but also offer a
unique opportunity for synergies between GW searches and CMB, large-scale structure, or even
dark-matter search experiments.

The detection of GW has opened up a new avenue for studying the early universe, which is complementary to other methods. Our analysis shows that cosmic strings can serve as excellent standard candles for probing the pre-BBN universe, should they exist 
International GW detector networks planned for the future could allow us to explore the equation of state of the universe down to $10^{-16}~\rm s$ after the Big Bang.

\section*{Acknowledgements}
YG and PS would like to express their gratitude to G{\'e}raldine Servant for fruitful discussions that initiated ideas explored in this paper and for providing valuable comments on the draft. We also appreciate the referee's useful remarks. YG thanks J{\'e}r{\^o}me Vandecasteele for stimulating interactions that contributed to this paper.
YG is grateful to the Azrieli Foundation for the award of an Azrieli Fellowship. PS is supported by Spanish AEI- MICINN, PID2020-113334GB-I00 /AEI/10.13039/501100011033 and the Generalitat Valenciana grant PROMETEO/2021/083.  LH acknowledges the support of the Science and Technology Facilities Council under Grant ST/T001011/1. LH acknowledges the support of the Institut Pascal at Université Paris-Saclay during the Paris-Saclay Astroparticle Symposium 2022, with the support of the IN2P3 master projet UCMN, the P2IO Laboratory of Excellence (program “Investissements d’avenir” ANR-11-IDEX-0003-01 Paris-Saclay and ANR-10-LABX-0038), the P2I axis of the Graduate School Physics of Université Paris-Saclay, as well as IJCLab, CEA, IPhT, APPEC,  and ANR-11-IDEX-0003-01 Paris-Saclay and ANR-10-LABX-0038. AG and PS appreciate the hospitality and partial support from the Mainz Institute for Theoretical Physics (MITP) of the Cluster of Excellence PRISMA$^+$ (Project ID 39083149)  during the early-stage work.

\appendix

\section{Effects from higher Fourier modes}
\label{app:higher_mode_effects}
\label{app:mode_max_perturbative}

\begin{figure}[!ht]
\centering
\includegraphics[width=0.485\textwidth, scale=1]{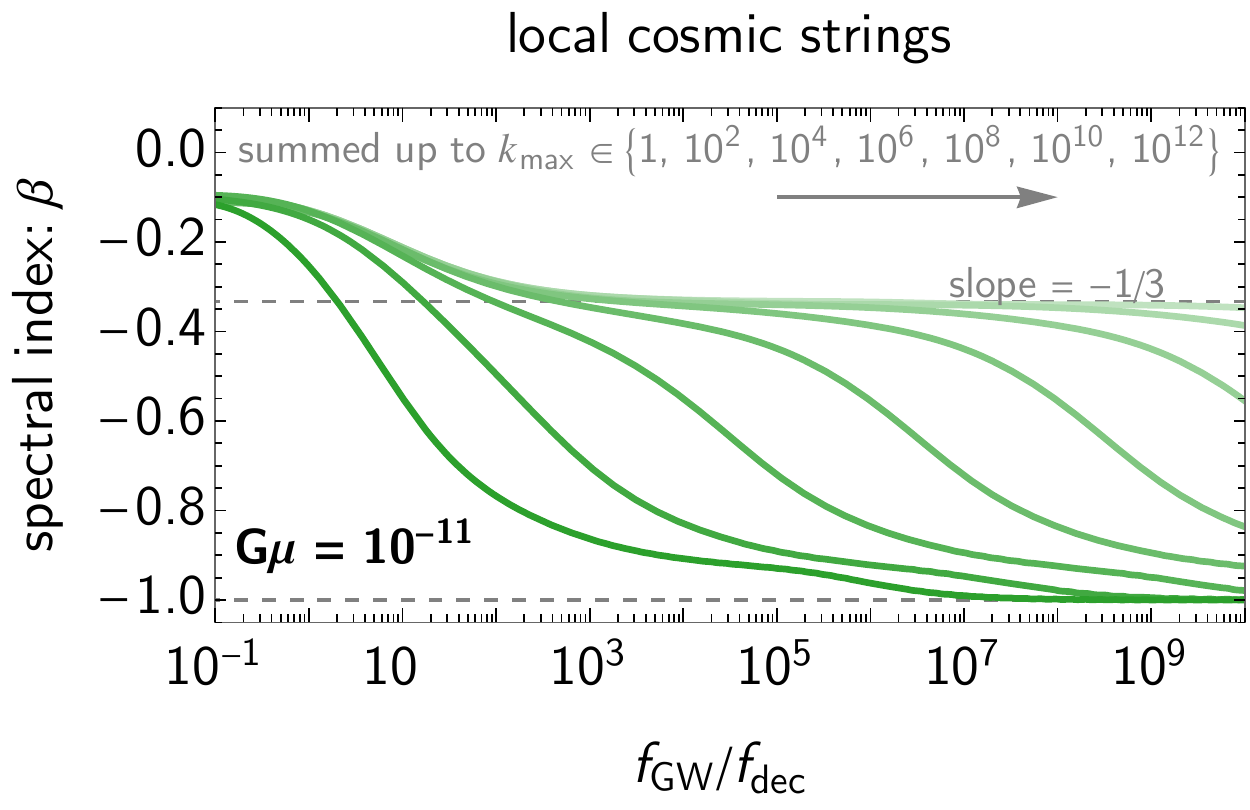}
\hfill
\includegraphics[width=0.485\textwidth, scale=1]{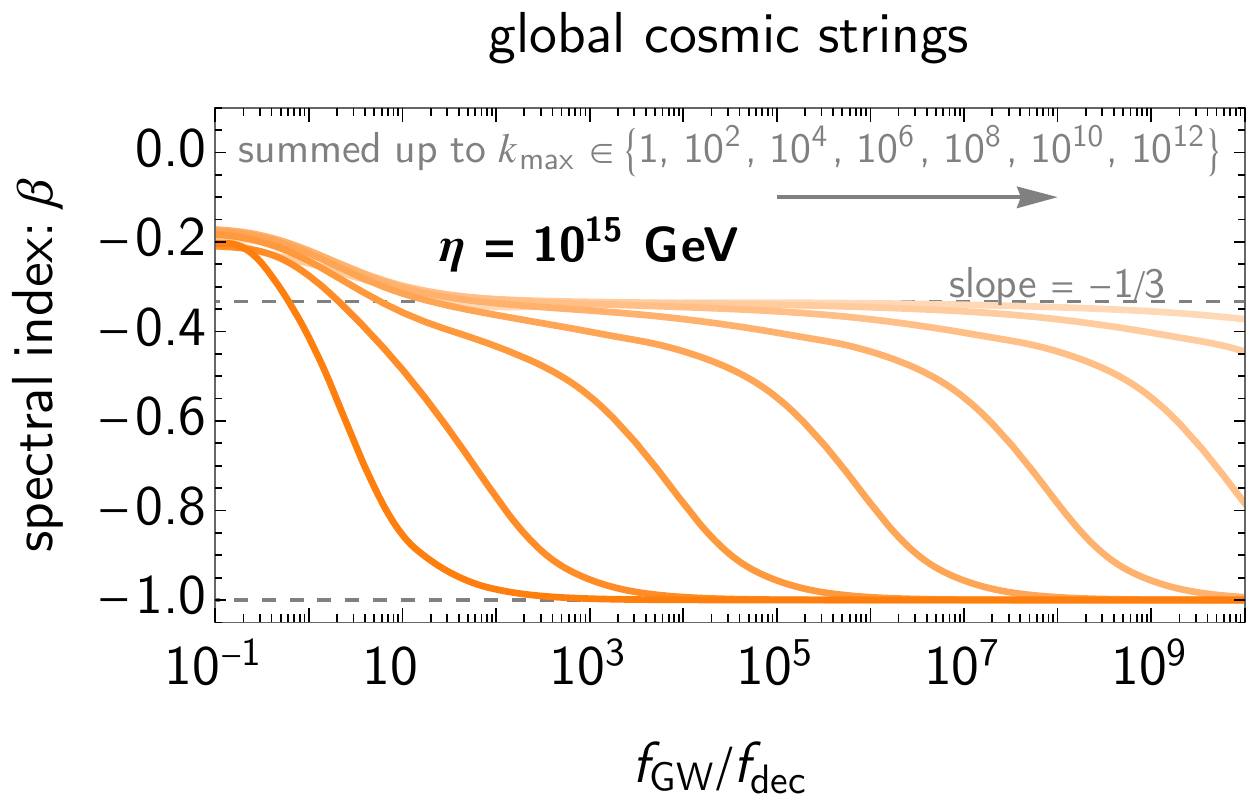}\\[-1em]
\caption{\textit{ \small The slopes of cosmic-string GW spectra assuming a long duration ($N_{\rm MD} = 30$) of EMD era ending at $T_{\rm dec} = 10^2 \, {\rm GeV}$. We can see that the first mode has spectral slope $f^{-1}$, while summation leads to $f^{-1/3}$ (up to the frequency $k_{\rm max} f_{\rm dec}$). }}
\label{fig:local_string_slope_summation}
\end{figure}

Due to periodic boundary conditions, any excitation of a cosmic string loop can be expressed as a superposition of Fourier modes with mode numbers $k$ ranging from 1 to a maximum value $k_{\rm max}$, which we now determine.
The GW spectrum given by Eq.~\eqref{eq:master_eq_pedagogical} assumes the Nambu-Goto (NG) approximation, which neglects the thickness of the string. This approximation breaks down when the string curvature becomes comparable to its thickness. The thickness of the string is estimated to be of the order of the inverse of the $U(1)$-breaking scalar vacuum expectation value $\eta$. Therefore, if the emitted GW frequency $\tilde{f}=2k/l$ becomes larger than $\eta$, the NG approximation is no longer valid. We deduce the highest Fourier mode that is compatible with the NG approximation (see also Ref.~\cite{Gouttenoire:2019kij})
\begin{align}
k_{\rm max} \simeq \frac{\eta L}{2} \simeq  \frac{\eta \alpha t}{2} \simeq \frac{\eta \alpha H^{-1}}{4} \simeq 4300 \left(\frac{\alpha}{0.1}\right) \left(\frac{10^{-10}}{G\mu}\right)^{\frac{1}{2}}\left(\frac{\eta}{T}\right)^2,
\end{align}
where $t$ is the time at which the loop is formed, $L\simeq \alpha t$ is the size of the loop, $H$ is the Hubble parameter, $g_*$ is the effective number of relativistic degrees of freedom, and $T\simeq 1.7\sqrt{M_{\rm pl}H}/g_*^{1/4}$ is the temperature of the radiation-dominated Universe. Trading the temperature $T$ of loop formation for the frequency $f$ of GW today, we obtain
\begin{align}
k_{\rm max} \simeq \left(\frac{\rm Hz}{f}\right)^2 \left[\frac{g_*(T(f))}{g_*(T_0)}\right]^{\frac{1}{2}} \times \begin{cases}
3.3 \cdot 10^{24} \left(\frac{10^{-11}}{G\mu}\right)^{1/2} \left(\frac{50}{\Gamma}\right) ~ &{\rm (local)},\\[0.5em]
2.3 \cdot 10^{21} \left(\frac{\eta}{10^{15} ~ {\rm GeV}}\right) \left(\frac{0.1}{\alpha}\right) ~ &{\rm (global)}.
\end{cases}
\label{eq:maxmodek}
\end{align}
The maximum mode number $k_{\rm max}$ is estimated to be extremely large, highlighting the importance of properly accounting for the contribution of each mode. In Fig.~\ref{fig:local_string_slope_summation}, we show effects from higher modes on the GW spectrum from CS, assuming a long EMD era.
 In this work, we account for those effects by summing over a large number of modes ($k_{\rm max} = 10^{12}$).

\section{High-frequency turning point for global strings}
\label{app:hf_turning_point_global}

\begin{figure}[!ht]
\centering
\includegraphics[width=\textwidth, scale=1]{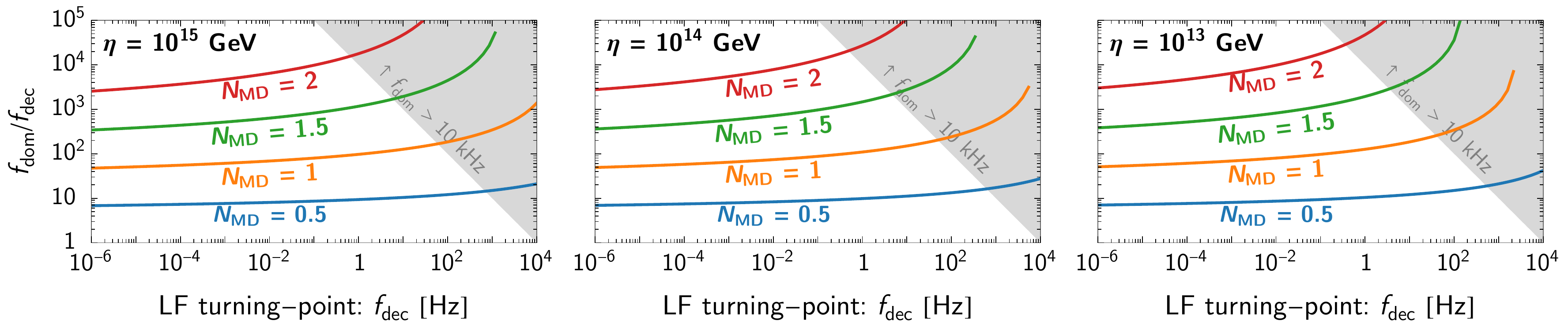}\\[-1em]
\caption{\textit{The HF turning-points $f_{\rm dom}$ (solid) of the global cosmic strings depends strongly on the $f_{\rm dec}$ and $N_{\rm MD}$. The log dependence shifts the turning point higher than the local case (dotted). In the gray region, the HF turning point sits at a frequency higher than $10$ kHz and in the ultra-higher frequency regime. This estimates well the numerical results in Fig.~\ref{fig:spectrum_strings}.
}}
\label{fig:hf_turning_point_global}
\end{figure}

As discussed in Sec.~\ref{sec:hf_plateau}, the EMD era leads to the global-string GW spectrum with the slope of $-1/3$. This leads to a high-frequency (HF) turning point (beyond which the HF plateau sits) at 
\begin{align}
f_{\rm dom} &\simeq f_{\rm dec} \left[\frac{\Omega_{\rm GW}(f_{\rm dec})}{\Omega_{\rm GW}(f_{\rm dom})}\right]^3
=  f_{\rm dec} \exp(3 N_{\rm MD})  \left[\frac{\Omega_{\rm GW}(f_{\rm dec})}{\Omega_{\rm GW}^{\rm st}(f_{\rm dom})}\right]^3,\nonumber\\
& \simeq  f_{\rm dec} \exp(3 N_{\rm MD}) 
\mathcal{D}^3(f_{\rm dec}, f_{\rm dom}),
\label{eq:turning_point_HF_glob}
\end{align}
\begin{align}
{\rm with} ~ ~ ~ ~ \mathcal{D}(f_{\rm dec}, f_{\rm dom}) \equiv \frac{\log^3\left[ (5.6 \times 10^{30}) \left(\frac{\eta}{10^{15} ~ {\rm GeV}}\right) \left( \frac{1 ~ {\rm mHz}}{f_{\rm dec}} \right)^2 \right]}{\log^3\left[ (5.6 \times 10^{30}) \left(\frac{\eta}{10^{15} ~ {\rm GeV}}\right) \left( \frac{1 ~ {\rm mHz}}{f_{\rm dom}} \right)^2 \right]}.\nonumber
\end{align}
The log-dependence term has the 9th power and strongly shifts the HF turning point.
Fig.~\ref{fig:hf_turning_point_global} shows that the HF turning point of global-string GW, defined in Eq.~\eqref{eq:turning_point_HF_glob}, sits many order-of-magnitude higher than the LF one, defined in Eq.~\eqref{turning_point_CS}. Consequently, the HF turning point may lie beyond the detectable range of forthcoming GW interferometers.

\section{More details on the \emph{knee} feature}
\label{app:knee_feature}

\subsection{Effects \emph{purely} from the early matter era (unrealistic GW spectrum)}
\label{app:spectrum_without_dof_evo}
\begin{figure}[!ht]
\centering
{\bf \small Gravitational waves from local cosmic strings}\\
{\small (unrealistic spectra assuming $g_*(T)=g_*(T_0)$ \& $g_{*s}(T)=g_{*s}(T_0)$)}\\
\centering
\includegraphics[width=0.485\textwidth, scale=1]{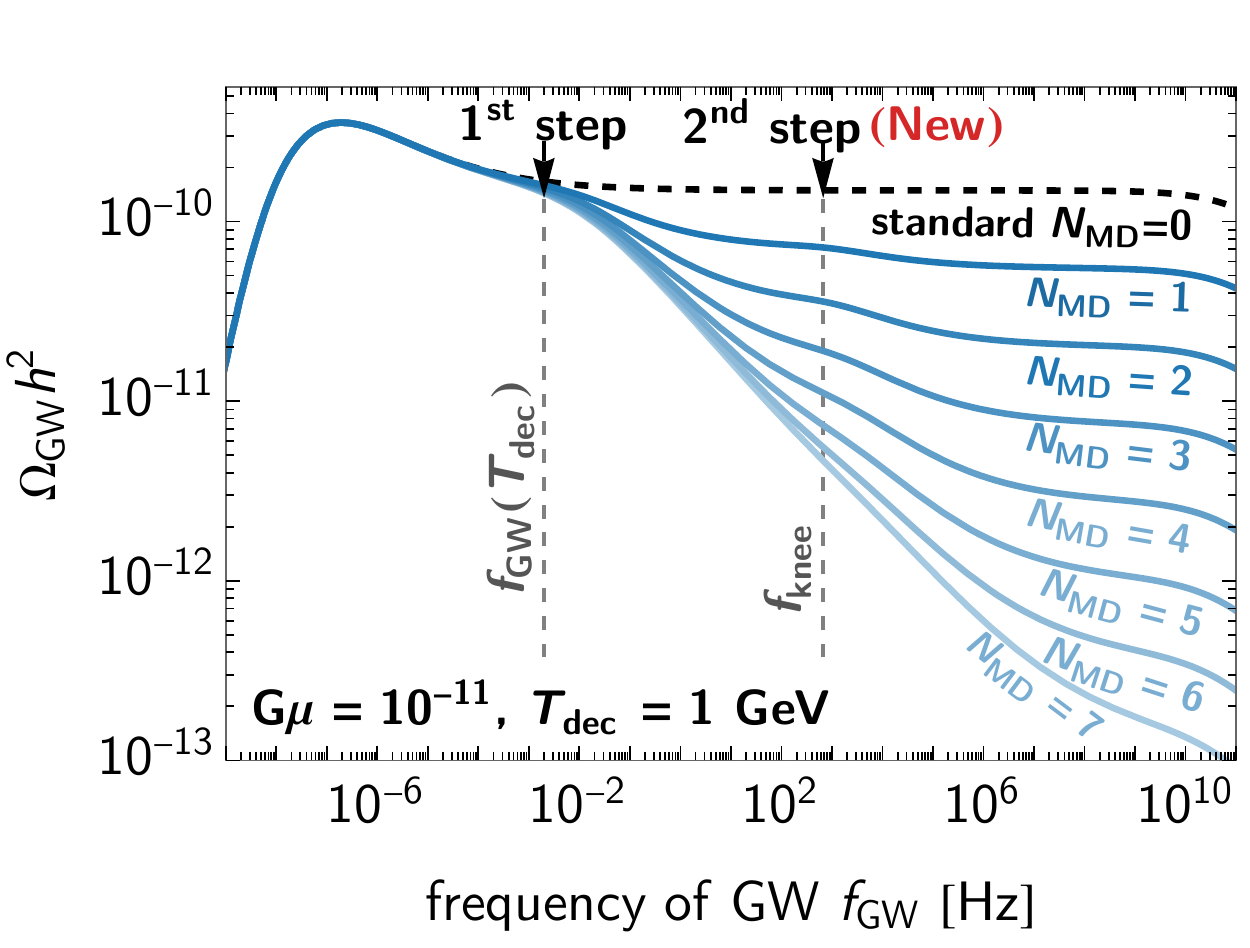}
\hfill
\includegraphics[width=0.485\textwidth, scale=1]{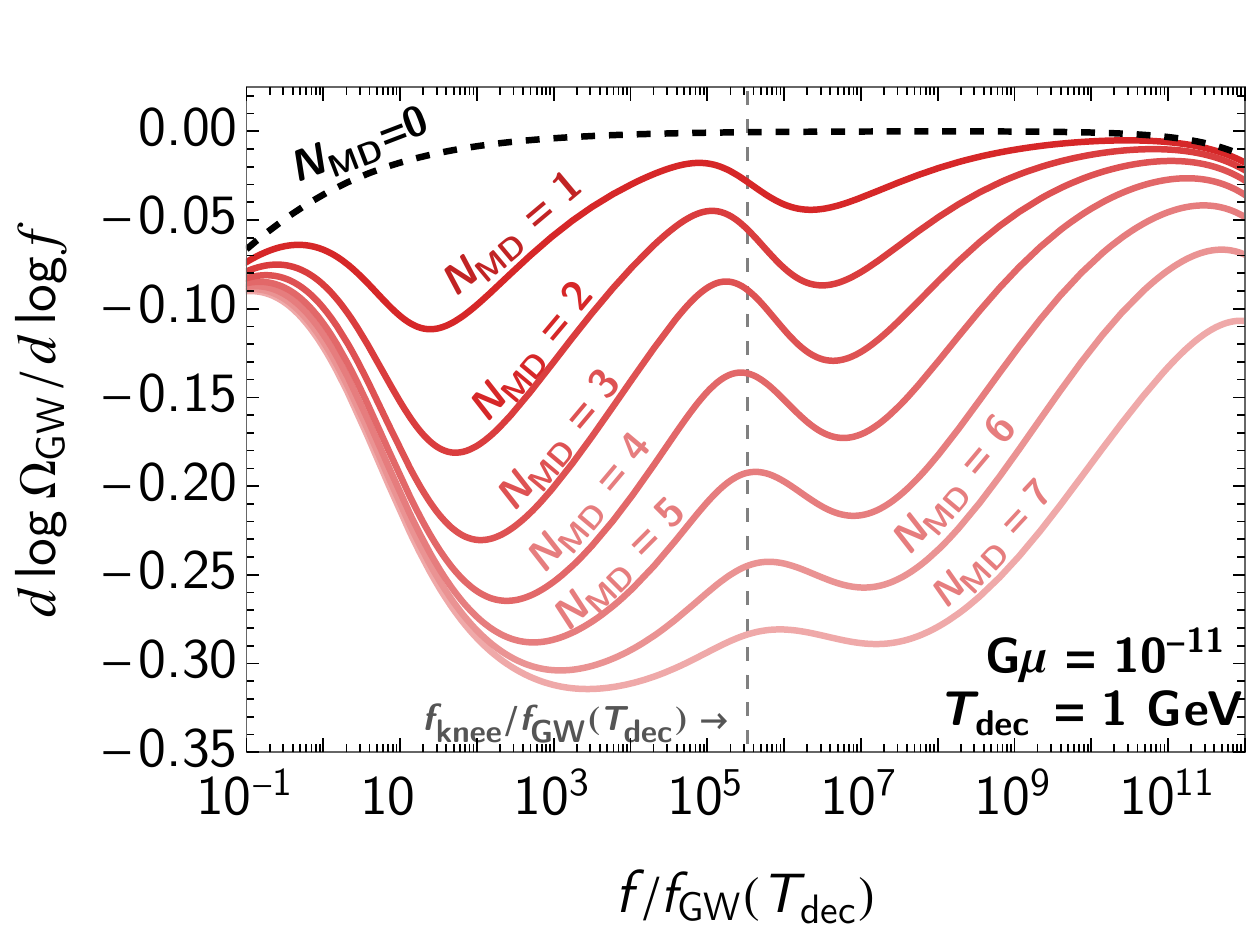}\\[-1em]
\caption{\textit{ \small The GW spectra (left) and their spectral indices (right) assume the early matter-domination era with $N_{\rm MD}$ e-folds. The amplitude of the high-frequency plateau and the position of the knee fit well with the analytic estimates in Eqs.~\eqref{eq:HF_plateau_amplitude}, \eqref{eq:knee_amplitude}, and \eqref{eq:knee_frequency} with $\mathcal{G} \to 1$.}}
\label{fig:spectrum_local_strings_no_dof}
\end{figure}

In Sec.~\ref{sec:matter_era}, we provide the analytic estimates for three features in the double-step GW spectrum from local strings. The $g_*$--$g_{*s}$ evolution contaminates the effects from the early matter era and complicates the comparison between Fig.~\ref{fig:spectrum_strings} and Eqs.~\eqref{eq:HF_plateau_amplitude}, \eqref{eq:knee_amplitude}, and \eqref{eq:knee_frequency}.
We now consider the effects purely from the early matter era by setting $g_*(T) = g_*(T_0)$ and $g_{*s}(T) = g_{*s}(T_0)$. We show the GW spectrum and its spectral indices in Fig.~\ref{fig:spectrum_local_strings_no_dof}. 
The features can be well described by our estimates in Eqs.~\eqref{eq:HF_plateau_amplitude}, \eqref{eq:knee_amplitude}, and \eqref{eq:knee_frequency}.
Furthermore, it can be inferred from Fig.~\ref{fig:spectrum_local_strings_no_dof} that the presence of double steps and knee features in the GW spectrum is not caused by the evolution of $g_*$ and $g_{*s}$, but is solely due to the early matter domination epoch.

\subsection{Visibility of the knee feature}
\label{app:knee_feature_visibility}
As discussed in Sec.~\ref{sec:knee_feature}, if the EMD era lasts too long, the knee feature disappears. This happens if the loop lifetime becomes shorter than the EMD duration, i.e., if 
\begin{align}
N_{\rm MD} > \log(\Tilde{a}_M/a_i) \simeq 12.28 + \frac{2}{3}\log\left[\left(\frac{\alpha}{0.1}\right)\left(\frac{50}{\Gamma}\right)\left(\frac{10^{-11}}{G\mu}\right)\right],
\end{align}
where we have used the lifetime of the loops in Eq.~\eqref{eq:tilde_t_M} and using $a_{\rm MD} \sim t^{2/3}$.

\section{Sensitivity curves of GW experiments}
\label{app:sensitivity_curve}

The sensitivity of a GW detector is $\Omega_{\rm sens} (f) \equiv 2 \pi^2 f^3 S_n(f)/(3 H_0^2)$, 
where $S_n(f)$ is the noise spectral density derived from the correlation of the detector noise signal $n(f)$: $\left<n^*(f)n(f')\right> \equiv \delta(f-f') S_n(f)$ \cite{Allen:1997ad,Maggiore:1999vm,Caprini:2018mtu}.
The ability of a GW detector to detect a GW signal with energy density $\Omega_{\rm GW}(f)$ is quantified by the signal-to-noise ratio (SNR),
\begin{align}
    {\rm SNR} \equiv \sqrt{T\int_{f_{\rm min}}^{f_{\rm max}} df \, \left[\frac{\Omega_{\rm GW}(f)}{\Omega_{\rm sens}(f)}\right]^2 } ~ ~ ~ ~ {\rm where} ~ T \equiv ~ {\rm an ~ observation ~ time}.
\end{align}
The calculation of the SNR can require expensive computations when scanning over the model parameter space. To avoid this, in this paper, we use \emph{the power-law integrated sensitivity curve} \cite{Thrane:2013oya}. 
We approximate the GW spectrum as a power-law $\Omega_{\rm GW}(f) = \Omega_\beta (f/f_{\rm ref})^\beta$ with a given spectral index $\beta$ and reference frequency $f_{\rm ref}$. We calculate the GW amplitude $\Omega_\beta$ which gives a certain SNR after an observation time $T$,
\begin{equation}
    \Omega_\beta =   \frac{\rm SNR}{\sqrt{T}}\left(\int_{f_{\rm min}}^{f_{\rm max}} df \, \left[\frac{(f/f_{\rm ref})^\beta}{\Omega_{\rm sens}(f)}\right]^2 \right)^{-1/2}.
\end{equation}
We now sample over all possible spectral index $\beta$.
One defines the envelope of these curves as the \emph{power-law integrated sensitivity curve},
\begin{align}
    \Omega_{\rm PI} (f;{\rm SNR},T) \equiv \max_\beta \left[f^\beta \frac{\rm SNR}{\sqrt{T}}\left(\int_{f_{\rm min}}^{f_{\rm max}} df \, \left[\frac{f^\beta}{\Omega_{\rm sens}(f)}\right]^2 \right)^{-1/2}\right].
\end{align}
For a detector with noise sensitivity $\Omega_{\rm sens}$, the observation during time $T$ of a GW signal with amplitude larger than $\Omega_{\rm PI}(f;{\rm SNR},T)$ has a signal-to-noise ratio  $>$ SNR.

The power-law integrated sensitivity curves $\Omega_{PI}(f)$, used in this study, are calculated from the noise spectral density in \cite{Hild:2010id} for ET, \cite{Evans:2016mbw} for CE, \cite{Yagi:2011wg} for BBO/DECIGO, \cite{AEDGE:2019nxb} for AEDGE, \cite{LISACosmologyWorkingGroup:2022jok} for LISA,  and \cite{Garcia-Bellido:2021zgu} for THEIA. We require SNR $= 10$ with the optimistic observation time of 10 years\footnote{The sensitivity is only lost by a small factor when the observation time is reduced by 10\%}.
For the sensitivity curves of pulsar timing arrays (EPTA, NANOGrav, and SKA), we directly took from \cite{Breitbach:2018ddu}.
The sensitivity curves of LIGO has been taken into account the improvement from the cross-correlation between multiple detectors \cite{Thrane:2013oya}, where we adopt the noise spectral densities for runs \href{https://dcc.ligo.org/LIGO-T1500293/public}{O2}, \href{https://dcc.ligo.org/LIGO-T1800044/public}{O4}, and \href{https://dcc.ligo.org/LIGO-T1800042-v4/public}{O5}, and the \href{https://dcc.ligo.org/public/0022/P1000128/026/figure1.dat}{overlap function} between the two LIGO detectors from \cite{LIGOScientific:2011yag}.
We fixed the LIGO curves at $\rm SNR = 10$ and the observational time of $T =$ 268 days for LIGO O2 and 1 year for LIGO O4 and O5.


\small
\bibliographystyle{JHEP}
\bibliography{refPBHCS.bib}

\end{document}